\documentclass[prb,showpacs,twocolumn,floatfix, groupaddress,superscriptaddress,footinbib]{revtex4-1}
\RequirePackage{silence}
\RequirePackage{float}
\setcitestyle{compressed}
\usepackage[utf8]{inputenc}
\usepackage[USenglish]{babel}
\usepackage[T1]{fontenc}
\usepackage{graphicx,epsf}
\usepackage[toc,page,header]{appendix}
\usepackage{minitoc}
\usepackage{subcaption}
\usepackage{color}
\usepackage{breakurl}
\usepackage[bookmarksnumbered,hypertexnames=false,colorlinks,colorlinks=true,linkcolor=blue,urlcolor=black,citecolor=blue,anchorcolor=green]{hyperref}
\usepackage{bbm,braket,microtype,mathrsfs,amsmath,amssymb,color,amsthm,mathtools,fullpage,enumitem,ae,aecompl,bm,thm-restate}
\usepackage{tikz}
\usepackage{lipsum}
\usepackage{nameref}
\usepackage{color}
\usepackage[capitalize]{cleveref}
\usepackage{silence}
\usepackage{mathtools}
\usepackage[font=small,justification=justified,format=plain]{caption}

\allowdisplaybreaks
\newcommand{\be}{\begin{equation}}
\newcommand{\ee}{\end{equation}}
\newcommand{\ba}{\begin{eqnarray}}

\newcommand{\ea}{\end{eqnarray}}
\DeclareMathOperator{\tr}{tr}

\newcommand{\ignore}[1]{}
\newcommand{\st}[1]{\ket{#1}\!\!\bra{#1}}

\newcommand{\pur}{\operatorname{Pur}}

\def\CC{{\rm\kern.24em \vrule width.04em height1.46ex depth-.07ex
   \kern-.29em C}}
\def\P{{\rm I\kern-.25em P}}
\def\RR{{\rm
        \vrule width.04em height1.58ex depth-.0ex
        \kern-.04em R}}
\def\bbbone{{\mathchoice {\rm 1\mskip-4mu l} {\rm 1\mskip-4mu l}
{\rm 1\mskip-4.5mu l} {\rm 1\mskip-5mu l}}}
\def\bbbc{{\mathchoice {\setbox0=\hbox{$\displaystyle\rm C$}\hbox{\hbox
to0pt{\kern0.4\wd0\vrule height0.9\ht0\hss}\box0}}
{\setbox0=\hbox{$\textstyle\rm C$}\hbox{\hbox
to0pt{\kern0.4\wd0\vrule height0.9\ht0\hss}\box0}}
{\setbox0=\hbox{$\scriptstyle\rm C$}\hbox{\hbox
to0pt{\kern0.4\wd0\vrule height0.9\ht0\hss}\box0}}
{\setbox0=\hbox{$\scriptscriptstyle\rm C$}\hbox{\hbox
to0pt{\kern0.4\wd0\vrule height0.9\ht0\hss}\box0}}}}
\def\bbbz{{\mathchoice {\hbox{$\sf\textstyle Z\kern-0.4em Z$}}
{\hbox{$\sf\textstyle Z\kern-0.4em Z$}}
{\hbox{$\sf\scriptstyle Z\kern-0.3em Z$}}
{\hbox{$\sf\scriptscriptstyle Z\kern-0.2em Z$}}}}

\newlength{\fighskip} \fighskip=2pt
\newlength{\figvskip} \figvskip=1pt

\makeatletter
\def\namedlabel#1#2{\begingroup
   \def\@currentlabel{#2}%
   \label{#1}\endgroup
}
\makeatother

\usepackage{hyperref}

\begin{document}
\setcounter{secnumdepth}{3}
%\onecolumngrid
\title{Stabilizer entropy dynamics  after a quantum quench}
\author{Davide Rattacaso}
\affiliation{Dipartimento di Fisica e Astronomia "G. Galilei", Universit\`a degli Studi di Padova, Via Marzolo 8, I-35131, Padova, Italy}
\affiliation{INFN, Sezione di Padova, Via Marzolo 8, I-35131, Padova, Italy}
\affiliation{Dipartimento di Fisica `Ettore Pancini', Universit\`a degli Studi di Napoli Federico II,
Via Cintia 80126,  Napoli, Italy}
\author{Lorenzo Leone}
\author{Salvatore F.E. Oliviero}
\affiliation{Physics Department,  University of Massachusetts Boston,  02125, USA}
\author{Alioscia Hamma}
\affiliation{Dipartimento di Fisica `Ettore Pancini', Universit\`a degli Studi di Napoli Federico II,
Via Cintia 80126,  Napoli, Italy}
\affiliation{INFN, Sezione di Napoli, Italy}
\begin{abstract}
 Stabilizer  entropies (SE) measure deviations from stabilizer resources and as such are a fundamental ingredient for quantum advantage. In particular, the interplay of SE and entanglement is at the root of the complexity of classically simulating quantum many-body systems. In this paper, we study the dynamics of SE in a quantum many-body system away from the equilibrium after a quantum quench in an integrable system. We obtain two main results: (i) we show that SE, despite being an $L$-extensive quantity, equilibrates in a time that scales at most linearly with the subsystem size; and (ii) we show that there is a SE length increasing linearly in time, akin to correlations and entanglement spreading.
\end{abstract}

\maketitle

\section{Introduction}

In the past few decades, significant progress in quantum information science has been closely linked to efforts to synthesize artificial many-body systems. These devices can be used to simulate the quantum dynamics of large systems and execute algorithms with a potentially exponential advantage over classical computers\cite{deutsch_rapid_1992,shor_algorithms_1994,lloyd_universal_1996,grover_fast_1996,kitaev_quantum_1997,abrams_quantum_1999,harrow_quantum_2009}. Benchmarking the effectiveness of these devices as quantum systems, in other words, understanding their quantumness, involves identifying the resources that hinder its classical simulation. Entanglement has been viewed as the necessary ingredient for quantumness~\cite{aspect1981experimental,aspect1982experimental,aspect1982experimental2} since the discovery of the Bell inequalities and the first experimental demonstrations~\cite{bell1964einstein,fine1982hidden,terhal2000bell,werner2001bell} and it plays  a fundamental role in the hardness of simulating of quantum many-body systems, e.g. in  Tensor Networks~\cite{latorre_ground_2004,orus_practical_2014}. 

Beyond entanglement, however, resources outside the stabilizer formalism\cite{gottesman1998HeisenbergRepresentationQuantum} are also necessary for  complex behavior in quantum many-body systems\cite{bravyi_universal_2005,campbell_bound_2010,campbell_unified_2017,howard_application_2017,seddon_quantifying_2019,leone_stabilizer_2022,leone_nonstabilizerness_2023, chamon_emergent_2014,yang_entanglement_2017,zhou_single_2020,liu_many-body_2022,white_conformal_2021,sewell_mana_2022,koukoulekidis_constraints_2022,hinsche_single_2022}.
%Magic in many-body systems
Recently,  stabilizer  entropy (SE)~\cite{leone_stabilizer_2022} has emerged as a measure of nonstabilizerness in quantum systems. Being an entropy, the SE can be moved around subsystems, with the effect of purifying those from nonstabilizerness, and can give rise to phase transitions as shown~\cite{leone_phase_2023,niroula_phase_2023}. Such quantity can be   experimentally measured on a quantum processor~\cite{oliviero2022MeasuringMagicQuantum,haug_scalable_2023} and its direct computability makes it amenable for the study of quantum many-body systems\cite{oliviero_magic-state_2022, haug_quantifying_2023,lami_quantum_2023,haug_stabilizer_2023, tirrito_quantifying_2023,odavic_complexity_2022,chen_magic_2022}.  

The locality of interactions implies that in the gapped ground state of one-dimensional systems, there is a finite correlation length $\xi$ such that the SE is localized\cite{oliviero_magic-state_2022,haug_quantifying_2023} within a  length $L_0\sim\xi$, in the sense that SE can be extrapolated by subsystems of size $L$ with an exponentially small error $\mathcal{O}(e^{-L/L_0})$. On the other hand, for critical systems $\xi$, is found to diverge resulting in a power law for the approximation error. 

In this paper, we investigate the behavior of SE in a quantum many-body system away from equilibrium after the quantum quench of an integrable spin chain. The time profile of SE is computed analytically for all times. The two main results of this paper are (i) SE equilibrates to the value of the infinite time average following a transient period that increases at most linearly with the size of the subsystem, and (ii) the SE length increases ballistically and is upper bounded by a spreading velocity that is proportional to the Lieb-Robinson speed for the system\cite{lieb2004finite}.

\begin{figure*}[t]
  \begin{subfigure}[t]{0.25\textwidth}
    \includegraphics[height=66mm]{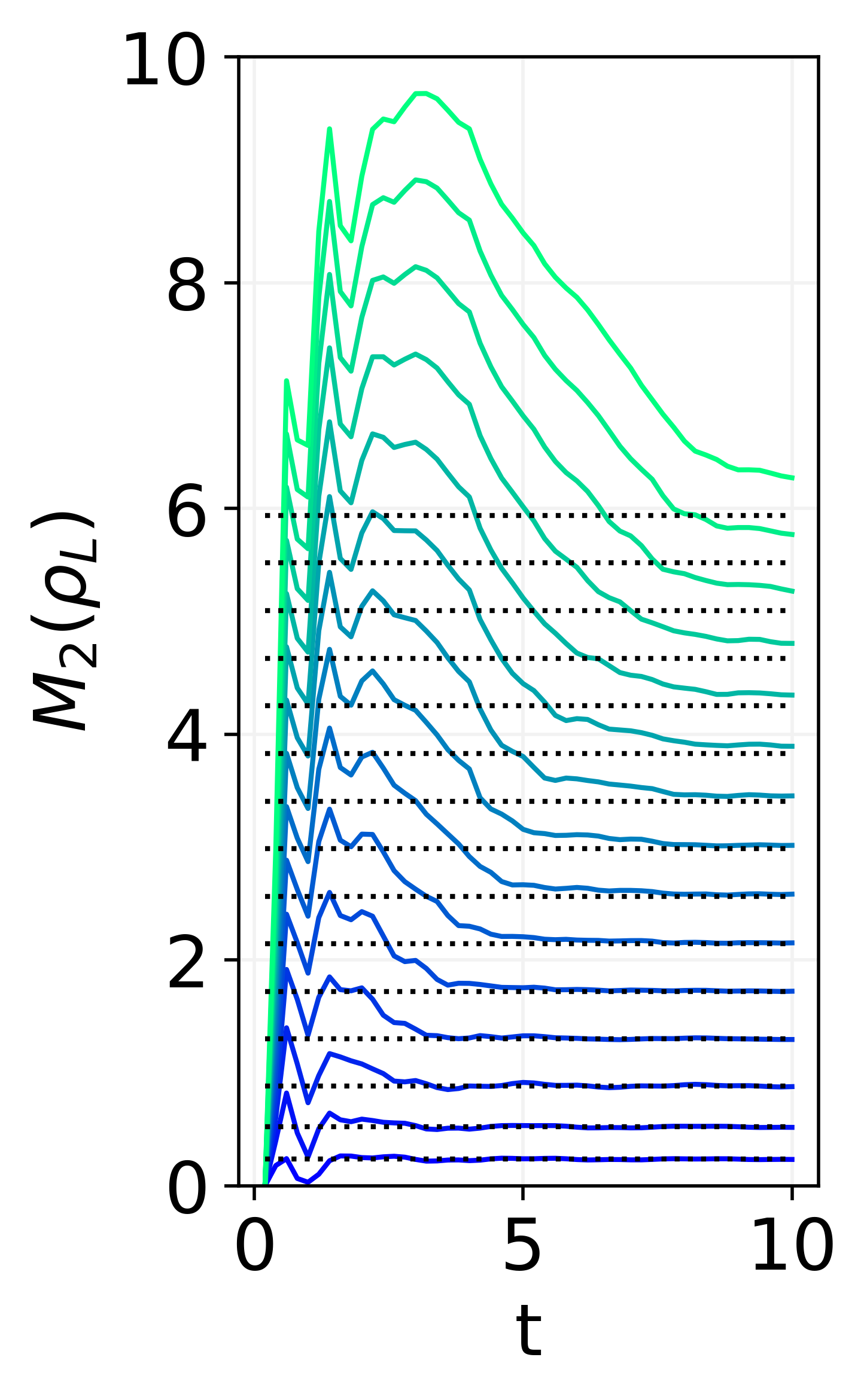}
    \caption{\kern-3em}
  \end{subfigure}\hfill
  \begin{subfigure}[t]{0.23\textwidth}
    \includegraphics[height=66mm]{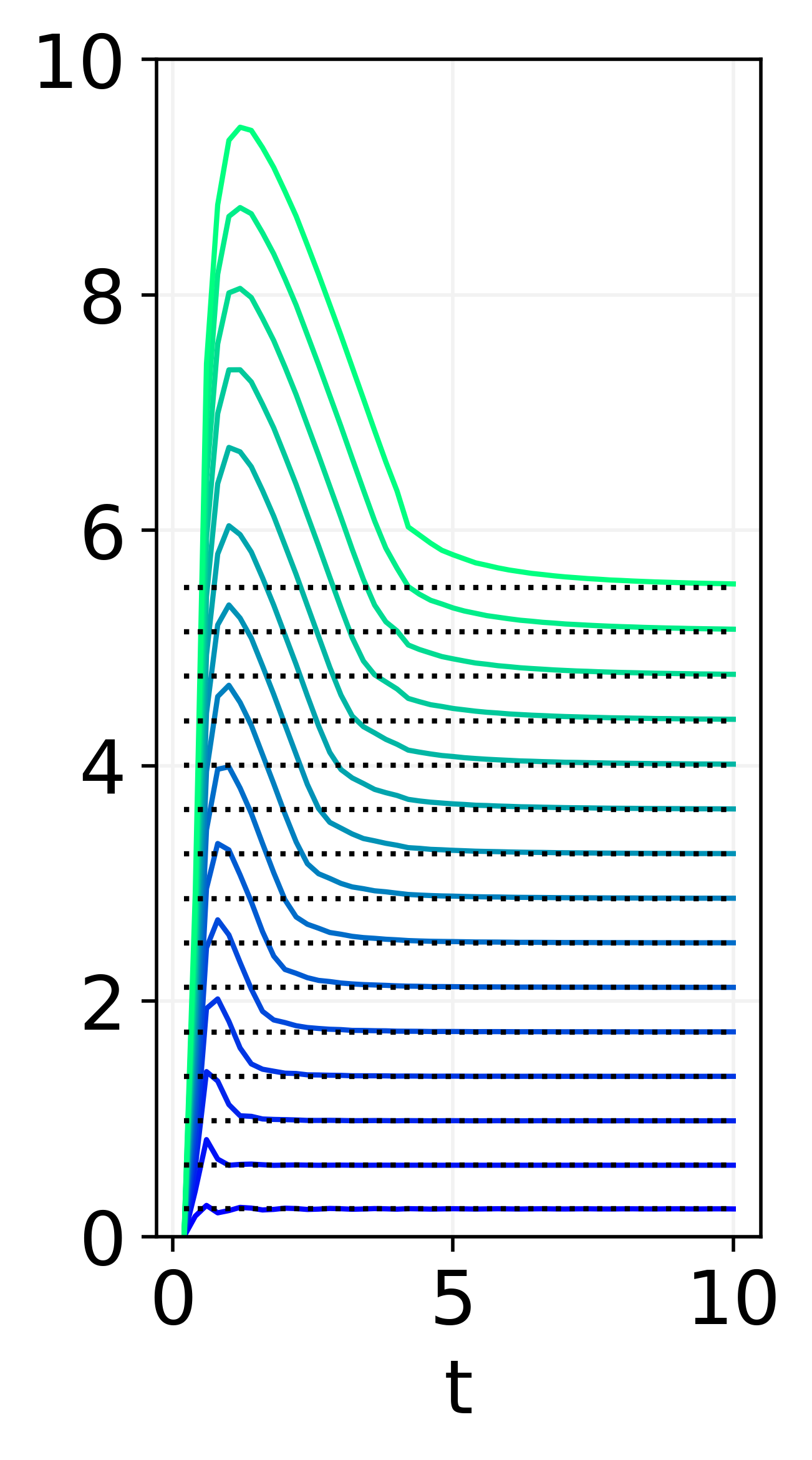}
    \caption{\kern-2em}
  \end{subfigure}\hfill
  \begin{subfigure}[t]{0.23\textwidth}
    \includegraphics[height=66mm]{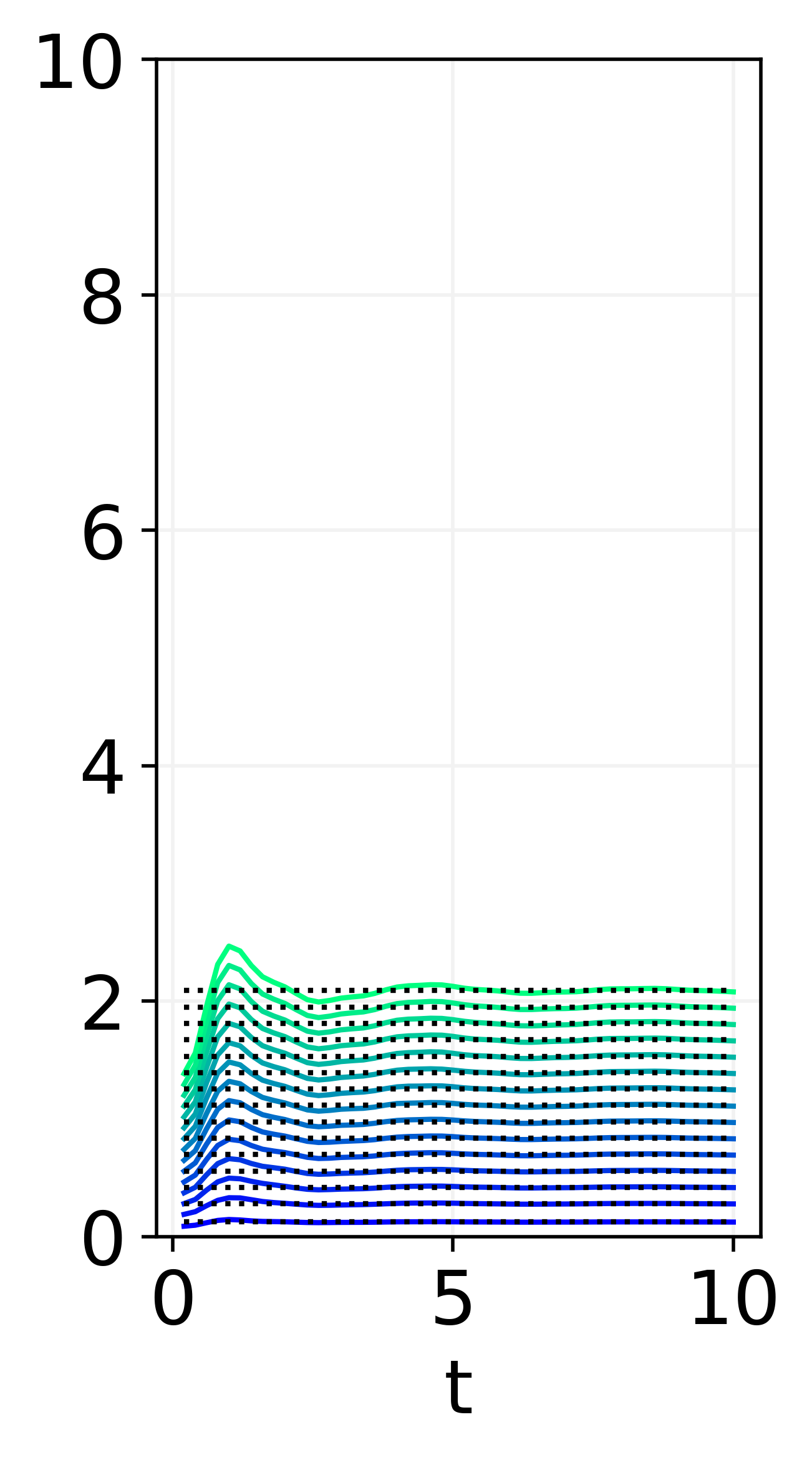}
    \caption{\kern-2em}
  \end{subfigure}\hfill
  \begin{subfigure}[t]{0.26\textwidth}
    \includegraphics[height=66mm]{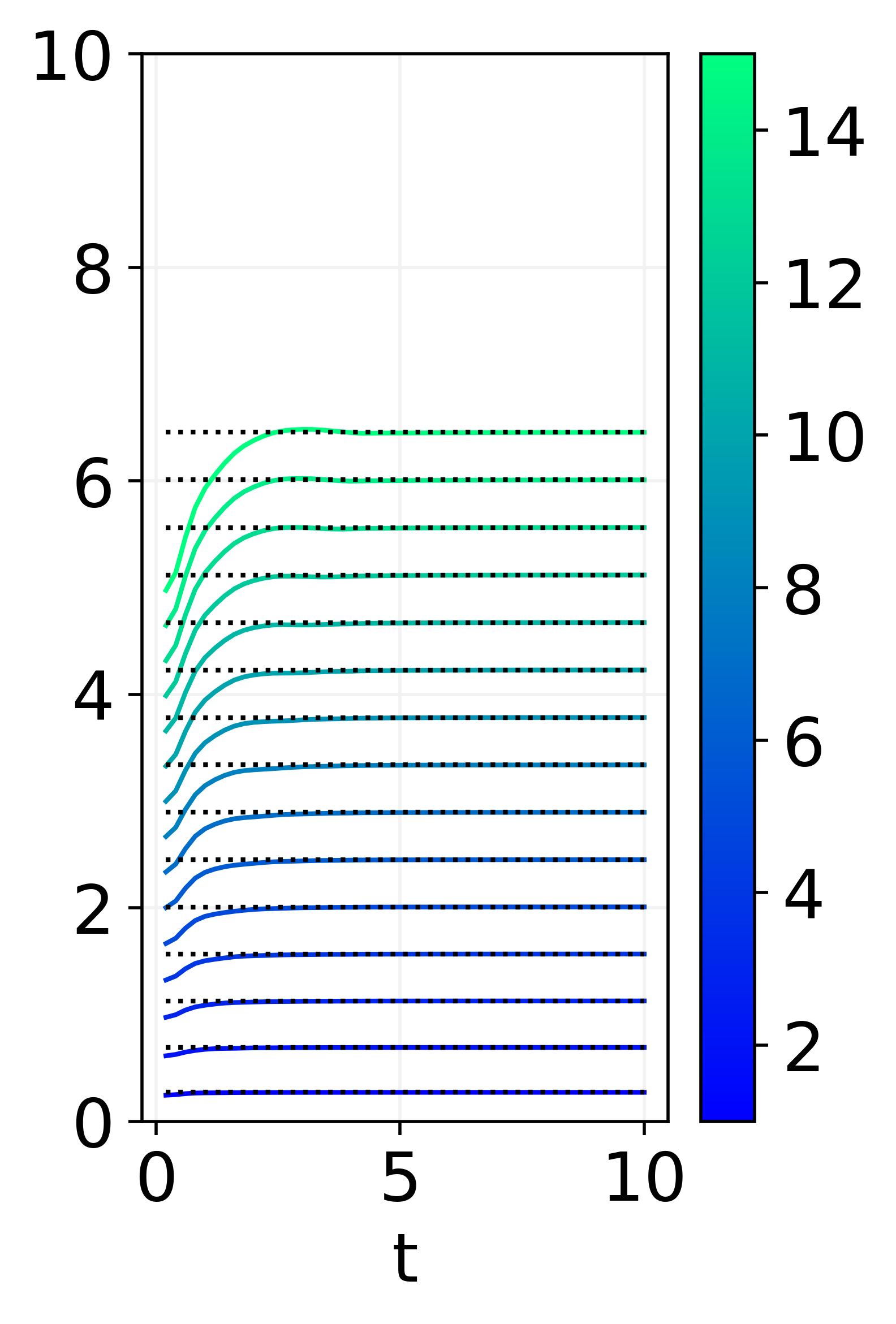}
    \caption{\kern-1em}
  \end{subfigure}
  \caption{\raggedright{Solid lines represent the time evolution of the SE $M_2$ (in Eq.~(\ref{eq:SE_def})) of connected subsystems of different sizes $L$, ranging from $1$ (blue lines below) to $15$ (green lines above), after a quench of the transverse field $\lambda$ in the Ising Hamiltonian in Eq.~(\ref{eq:ising}). Dotted black lines represent the SE of the corresponding dephased subsystems states $\overline{\psi}_L=\tr_{N/L} \lim_{T\to\infty}T^{-1}\int_0^T\psi_t dt$. In Panel (a) for the large quench $\lambda:10^4\rightarrow 0.5$; in Panel (b) for the large critical quench $\lambda:10^4\rightarrow 1.0$; in Panel (c) for the small quench $\lambda:0.5\rightarrow 0.6$; in Panel (d) for the small critical quench $\lambda:0.9\rightarrow 1.0$. After a time transient, the SE equilibrates to the dephased state value, which is upper-bounded by $L/2$ as expected for an integrable dynamic.}}
    \label{fig:M2_time}
\end{figure*}

The paper is organized as follows. First, in Section~\ref{sec:SE_away} we introduce the stabilizer entropy (SE) of a subsystem and investigate its evolution in the 1-dimensional transverse field Ising model after a quench. We show that the SE equilibrates to the infinite time average and, after a large quench, the equilibration time of subsystem of size $L$ scales as $L/v_{LR}$, where $v_{LR}$ is the Lieb-Robinson speed associated to the quench. In Section~\ref{sec:SE_length} we introduce the notion of SE length. Looking at its time evolution, we can investigate how non-stabilizerness dynamically delocalizes: first, in Subsection~\ref{subsec:SE_deloc}, we use an analytical argument to show that the SE delocalizes in a light-cone, then in Subsection~\ref{subsec:SE_deloc_TFIM} we show that in the TFIM this length grows ballistically with a speed proportional to the Lieb-Robinson velocity. Finally, Section~\ref{sec:conclusions} is devoted to conclusions and future perspectives.

\section{SE away from equilibrium}\label{sec:SE_away}

Let us start with the definition of SE\cite{leone_nonstabilizerness_2023}. Let  ${\psi}$ be a pure state  of a system with $N$ qubits on a chain and  $\rho_L:=\tr_{N\setminus L}\psi$ its reduced density operator to a subsystem of $L$ contiguous qubits. 
Denote $\mathcal{P}_L$  the Pauli group on such a subsystem. The SE  (of order two) of $\rho_L$ is defined as
\begin{equation}\label{eq:SE_def}
    M_2(\rho_L):=-\log_2W(\rho_L)-S_{2}(\rho_L)
\end{equation}
where $W(\rho_L):=2^{-L}\sum_{P\in\mathcal{P}_L}\tr[(P\rho_L)^4]$ is the so-called \textit{stabilizer purity} of $\rho_L$, while $S_{2}=-\log\operatorname{Pur}(\rho_L)$ is $2$-R\'enyi entanglement entropy of $\rho_L$ and $\operatorname{Pur}(\rho_L)\equiv\tr(\rho_L^2)$. SE is a good measure of nonstabilizerness from the point of view of resource theory. Indeed, it has the following properties: (i) faithfulness $M_{\alpha}\left(|\psi\rangle \right)=0$ iff $|\psi \rangle \in \rm{STAB}$, otherwise $M_{\alpha}(|\psi\rangle) >0$, (ii) stability under Clifford operations: $\forall\,\Gamma  \in \mathcal{C}_n$ we have that $M_{\alpha} \left( \Gamma |\psi \rangle \right) = M_{\alpha} \left(|\psi\rangle \right)$ and (iii) additivity  $M_{\alpha} \left( |\psi \rangle \otimes |\phi \rangle \right) = M_{\alpha} \left( |\psi\rangle \right) + M_{\alpha} \left( |\phi\rangle \right) $ (the proof can be found in \cite{leone_nonstabilizerness_2023}). However, the SE with R\'enyi index $0<\alpha<2$ are shown to be non-monotone under measurements followed by conditioned Clifford transformations, see~\cite{haug_stabilizer_2023}.  %is a measure of the spreading of the state decomposition on the basis of Pauli strings. Indeed, identifying the probability distribution in the Pauli basis $\mathcal{P}_L$ with components $\Xi_{\rho_L}(P)=\operatorname{Pur}^{-1}(\rho_L)d^{-1}\tr^2(P\rho_L)$, its purity $M_{2}(\rho_L)\propto\sum_{P}\Xi_{\rho_L}^2(P)$ quantifies the spreading of $\rho_L$ in the Pauli basis $\mathcal{P}_L$.

Let us now turn our attention to the paradigmatic example of a family of $1$-parameter Hamiltonians: the $1$-dimensional transverse-field Ising model (TFIM) $H_I(\lambda)$ defined as 
\begin{equation}\label{eq:ising}
H_I(\lambda)=-\sum_n\left(\sigma_n^x\sigma_{n+1}^x+\lambda\sigma_n^z\right)  
\end{equation}
with periodic boundary conditions $\sigma_{N+1}:= \sigma_1$. In Eq.~\eqref{eq:ising} $\sigma_{i}^{\alpha}$ for $\alpha=x,y,z$ are Pauli matrices acting on the $i$-th spin and $\lambda$ is the strength of the transverse field. The model is integrable for any value of $\lambda$ using the Jordan-Wigner transformation and the Wick theorem~\cite{lieb_two_1961,PFEUTY197079,barouch_statistical_1971}. Denote $\ket{\psi_{0}(\lambda)}$ the ground state of $H(\lambda)$.  To study the dynamics of SE we subject the system to a  quantum quench  $H_I(\lambda)\mapsto H_I(\lambda^\prime)$ and let $\ket{\psi_0(\lambda)}$ evolve under the unitary evolution generated by $H(\lambda^\prime)$
\be
\ket{\psi_0(\lambda)}\mapsto\ket{\psi_t(\lambda,\lambda')}=e^{-iH(\lambda^{\prime})t}\ket{\psi_0(\lambda)}
\label{quencheq}
\ee
Thanks to Wick's theorem, the time-dependent expectation values $P_t\equiv \tr[P\psi_t(\lambda,\lambda')]$ can be computed analytically for any subsystem of length $L$ in the thermodynamic limit $N\to\infty$~\cite{IsingSantoro}. However, since there are $4^N$ such expectation values, we evaluate them for subsystems of sizes $L=1,\ldots, 16$.

In Fig.\ref{fig:M2_time} we show the evolution of the SE of subsystems under different quenches $H(\lambda')$, and the SE $M_2(\overline{\rho}_L)$ of the dephased state 
$\overline{\rho}_L\equiv\tr_{N/L} \sum_k \Pi_k(\lambda') \psi_0(\lambda)\Pi_k(\lambda')=\tr_{N/L} \lim_{T\to\infty}T^{-1}\int_0^T\psi_t dt$. This is the infinite time average of the state, corresponding to the completely dephased state in the basis $\{\Pi_k(\lambda')\}$ of the Hamiltonian $H_I(\lambda')$\cite{rigol_thermalization_2008,eisert_quantum_2015}. The initial state is chosen to be the completely polarized state for $\lambda\gg 1$, which is a stabilizer state so that $M_2(t=0) =0$. As we can see, there is a transient in which $M_2(t)$ increases rapidly before equilibrating to the SE of the dephased state. The equilibration is noteworthy because $M_2$ is an $L$-extensive quantity so it is not assured to equilibrate at finite $t$ for every size $L$ under general conditions\cite{PhySEvLett.80.1373, PhySEvLett.101.190403, PhySEvE.79.061103}.

\begin{figure}[t]
\centering  
  \begin{subfigure}[b]{\linewidth}
  \hspace*{-0.7cm}
    \includegraphics[width=\linewidth]{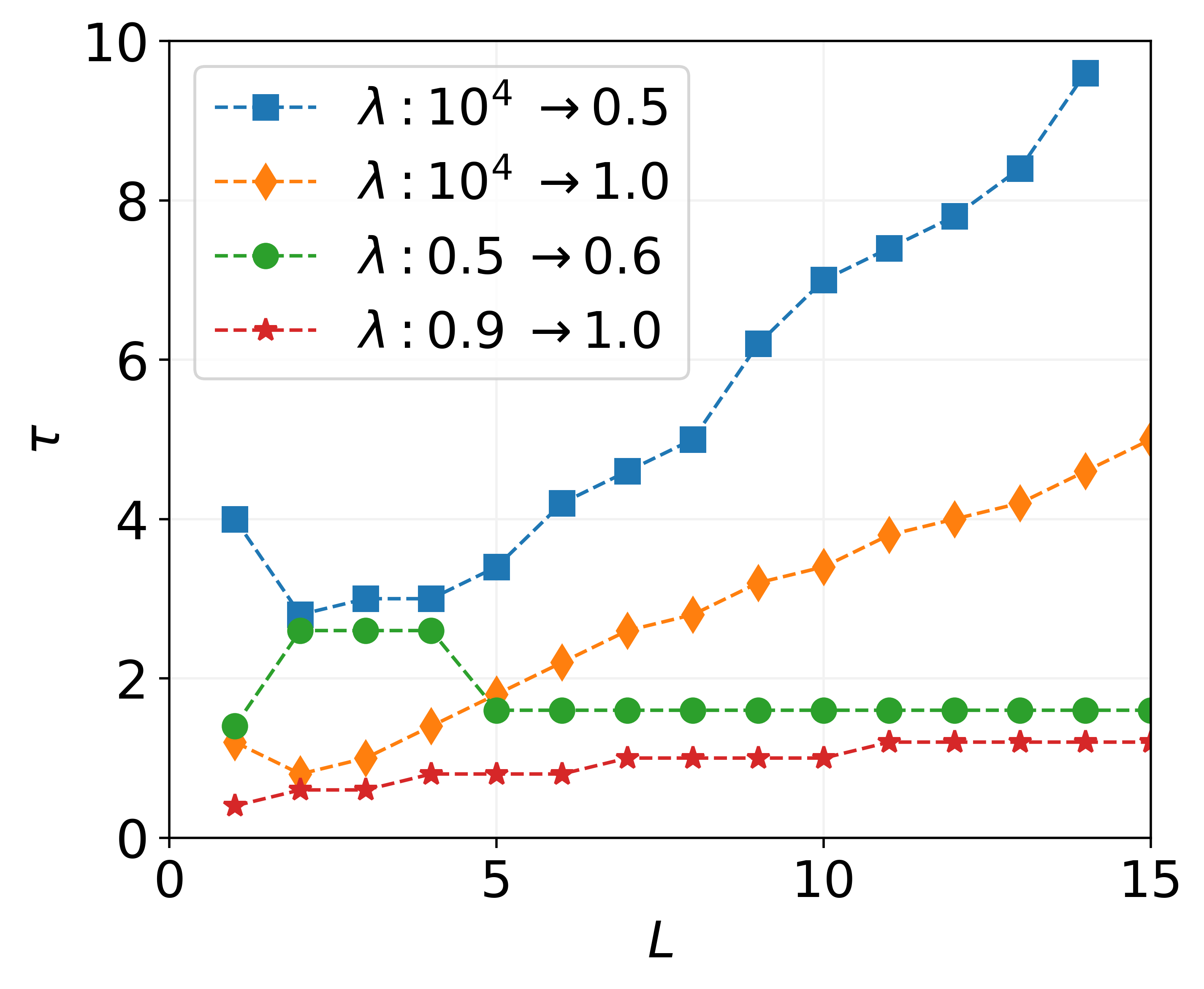}
    \caption{\kern-1em}
    \hspace*{0.7cm}
  \end{subfigure}\hfill
  \begin{subfigure}[b]{\linewidth}
    \hspace*{-0.7cm}
\includegraphics[width=\linewidth]{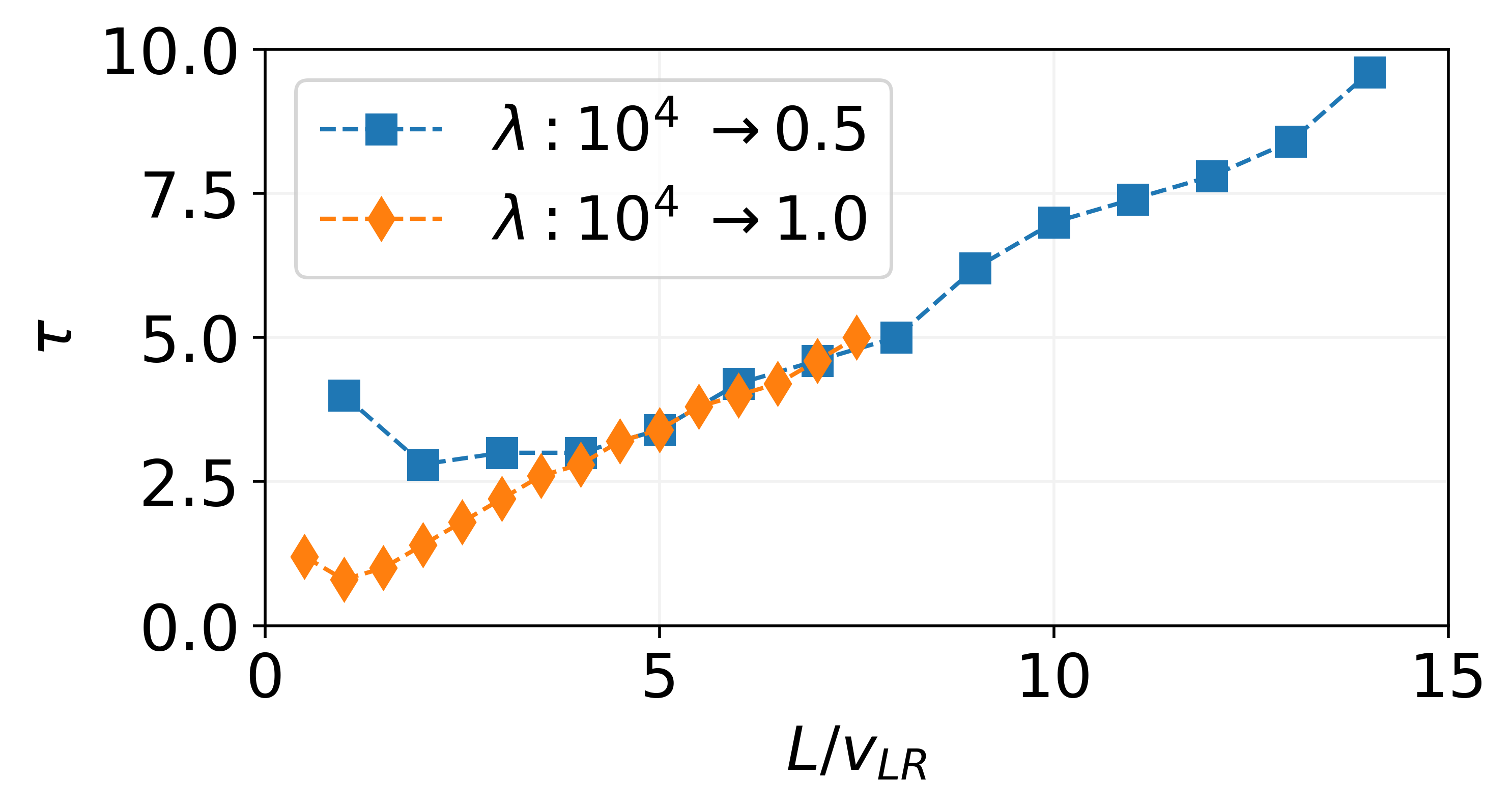}
    \caption{\kern-1em}
  \hspace*{0.7cm}
  \end{subfigure}
  \caption{\raggedright{In Panel (a), the equilibration time ($\tau$) for the subsystem's SE as a function of the number of sites ($L$) ranging from $0$ to $15$. Here the equilibration time is defined as the time it takes for the subsystem's SE to reach the SE of the dephased state with a tolerance of $5\%$. We consider four types of quenches: a large non-critical quench (represented by blue squares), a large critical quench (represented by yellow diamonds), a small non-critical quench (represented by green dots), and a small critical quench (represented by red stars). In panel (b) the number of sites is rescaled by the Lieb-Robinson speed. We can see} that the equilibration time after a large quench scales as $\tau\simeq L/v$, where $v_{LR}$ is the Lieb-Robinson speed of signals associated with the quench Hamiltonian and $v\propto v_{LR}$.}
    \label{fig:eq_time}
\end{figure}

For each configuration, the equilibration time $\tau$ can be defined as the time it takes for the subsystem's SE to reach the SE of the dephased state with fixed tolerance. The scaling of $\tau$ as a function of the system size is depicted in Figure~\ref{fig:eq_time} for different quenches. Here we can see that, after a large quench, the equilibration time can be computed as $\tau \simeq L/v$ with $v\propto v_{LR}$, the Lieb-Robinson speed of propagation of signals reconstructed in Appendix~\ref{app:lr_speed}.

\section{SE length dynamics}\label{sec:SE_length}

The localization of SE in a quantum many-body system is described by the SE length. Localization of SE makes this quantity more amenable to computation in large systems. Indeed, although its computation does not involve a minimization procedure\cite{campbell_bound_2010,campbell2017roads,seddon_quantifying_2019}, it is  still exponentially expensive as the number of Pauli operators $P$ is $4^N$, as we have seen in the previous section.

\begin{figure*}[t]
  \begin{subfigure}[t]{0.27\textwidth}
    \includegraphics[height=66.5mm]{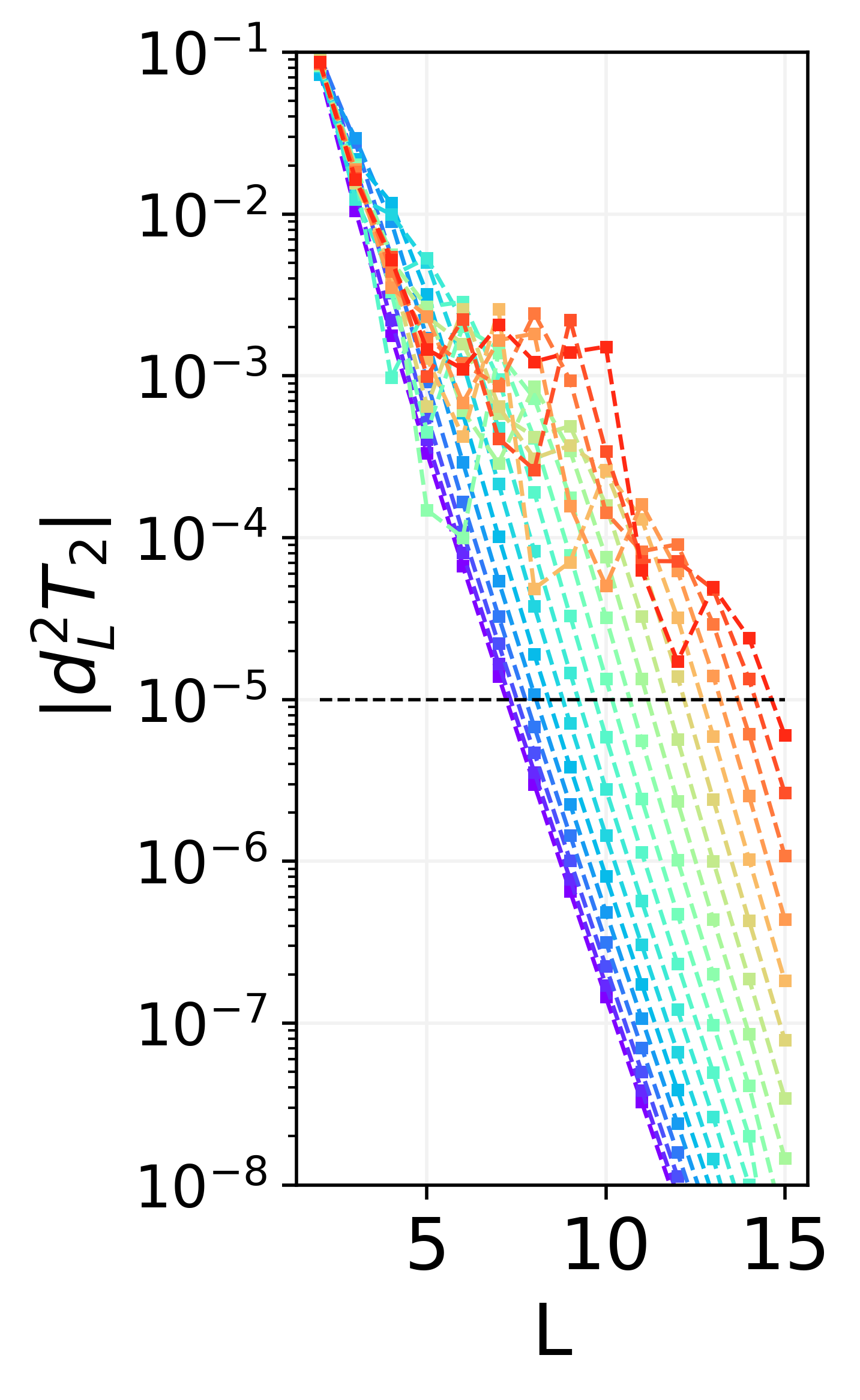}
    \caption{\kern-4em}
  \end{subfigure}\hfill
  \begin{subfigure}[t]{0.23\textwidth}
    \includegraphics[height=66mm]{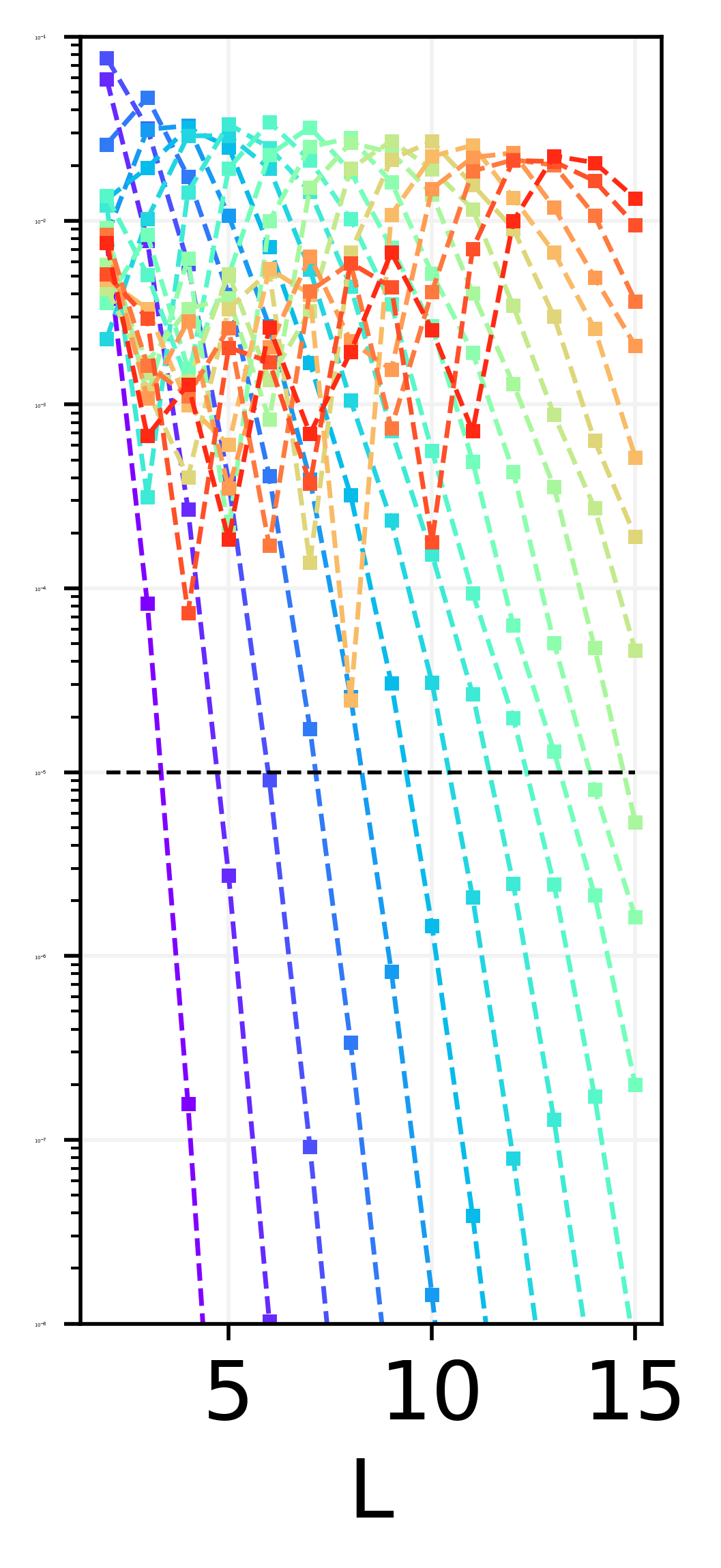}
    \caption{\kern-1em}
  \end{subfigure}\hfill
  \begin{subfigure}[t]{0.23\textwidth}
    \includegraphics[height=66mm]{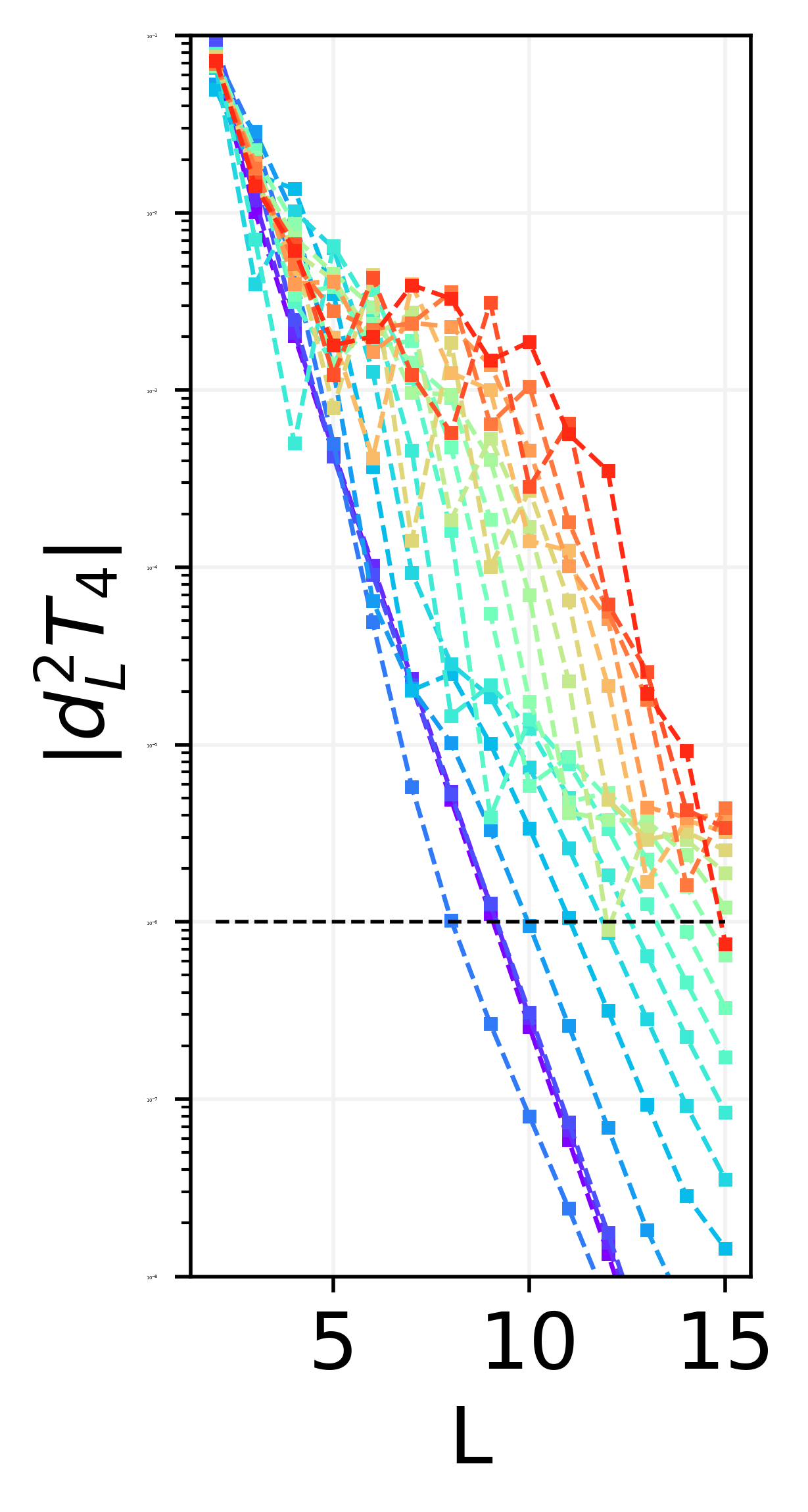}
    \caption{\kern-2em}
  \end{subfigure}\hfill
  \begin{subfigure}[t]{0.27\textwidth}
    \includegraphics[height=66.5mm]{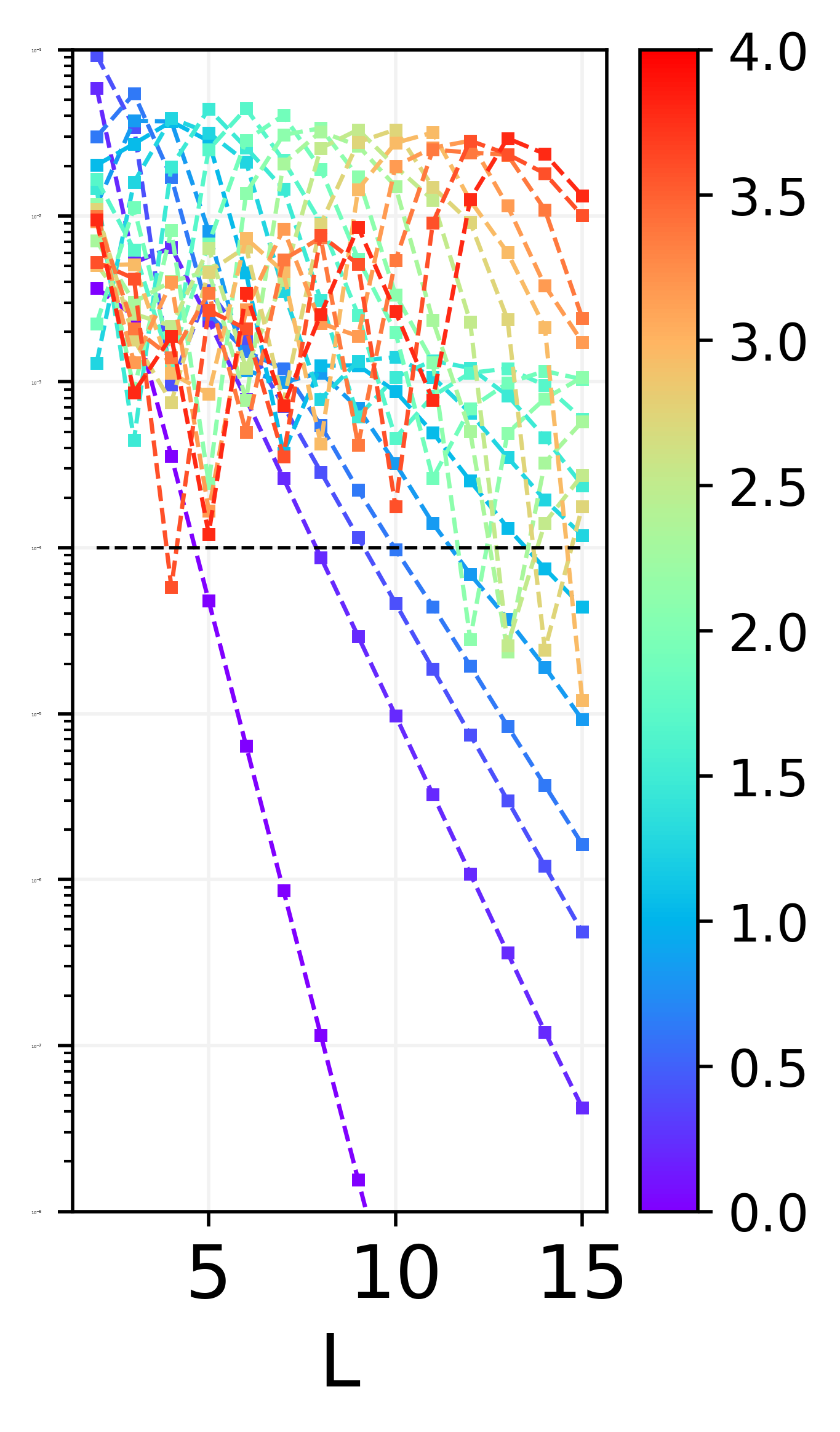}
    \caption{\kern+1em}
  \end{subfigure}
  \caption{\raggedright{Second space derivative of $T_2(\rho_L)$ and $T_4(\rho_L)$ (see Eq.~(\ref{secondderivative})) as a function of the subsystem size $L$, ranging from $2$ to $15$, after the quench. Different lines represent different times in the range from $0$ (purple line below) to $4$ (red line above). The horizontal black line represents the error tolerance such that functions above the line decay exponentially. In Panel (a) second derivative of $T_2$ for the small quench $\lambda:0.5\rightarrow 0.6$. In Panel (b) second derivative of $T_2$ for the large critical quench $\lambda:10^4\rightarrow 1.0$. In Panel (c) second derivative of $T_4$ for the small quench $\lambda:0.5\rightarrow 0.6$. In Panel (d) second derivative of $T_4$ for the large critical quench $\lambda:10^4\rightarrow 1.0$. All these derivatives can be upper-bounded by exponentially decaying functions: the SE additivity in Eq.~(\ref{additivemagic}) is preserved in time evolution with an exponentially small error for sufficient large subsystems.}}
    \label{fig:ddT}
\end{figure*}

\begin{figure}[t]
  \begin{subfigure}[b]{\linewidth}
  \hspace*{-0.7cm}
    \includegraphics[height=51mm]{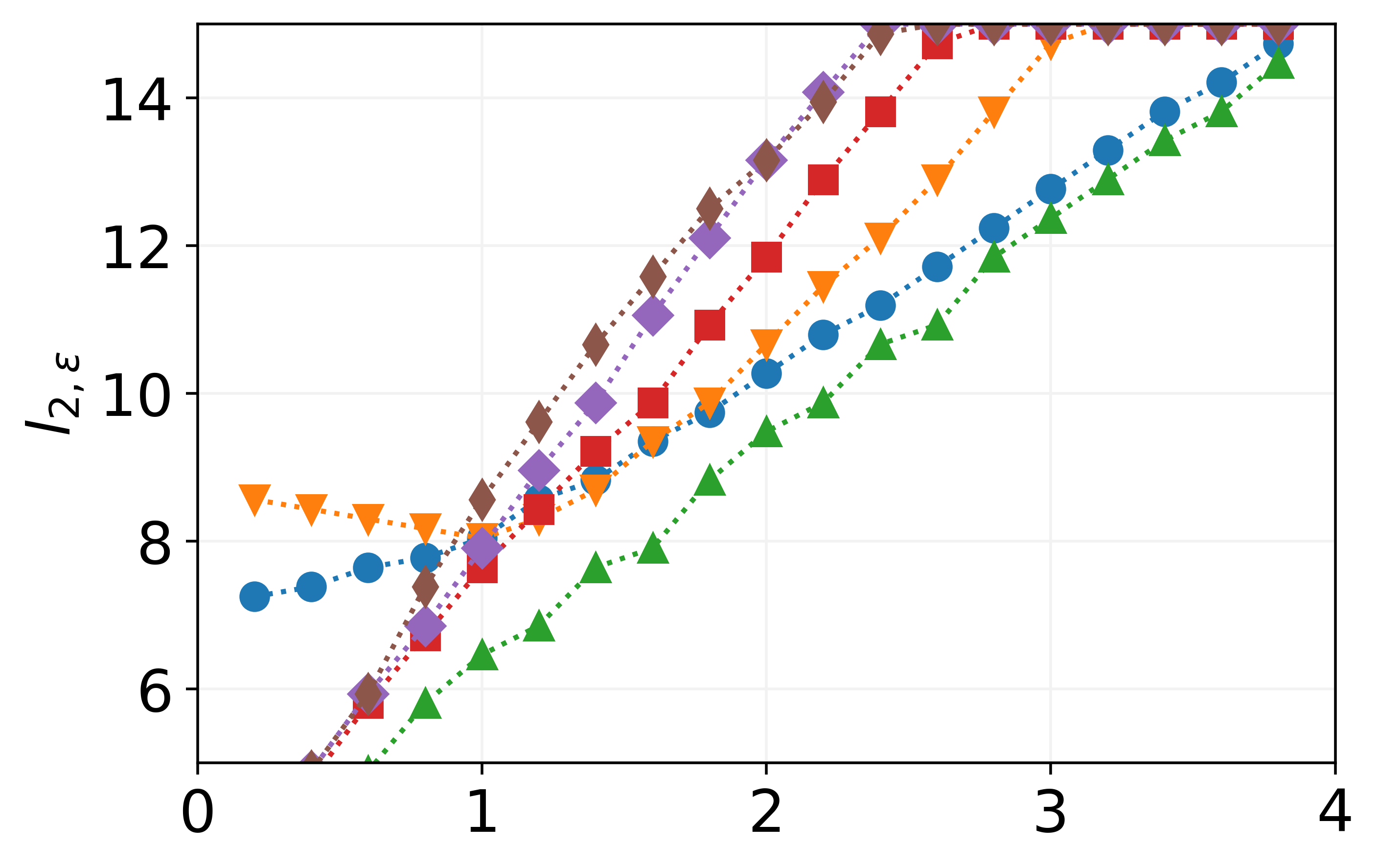}
    \hspace*{0.7cm}
  \end{subfigure}\hfill
  \begin{subfigure}[b]{\linewidth}
    \hspace*{-0.7cm}
\includegraphics[height=55mm]{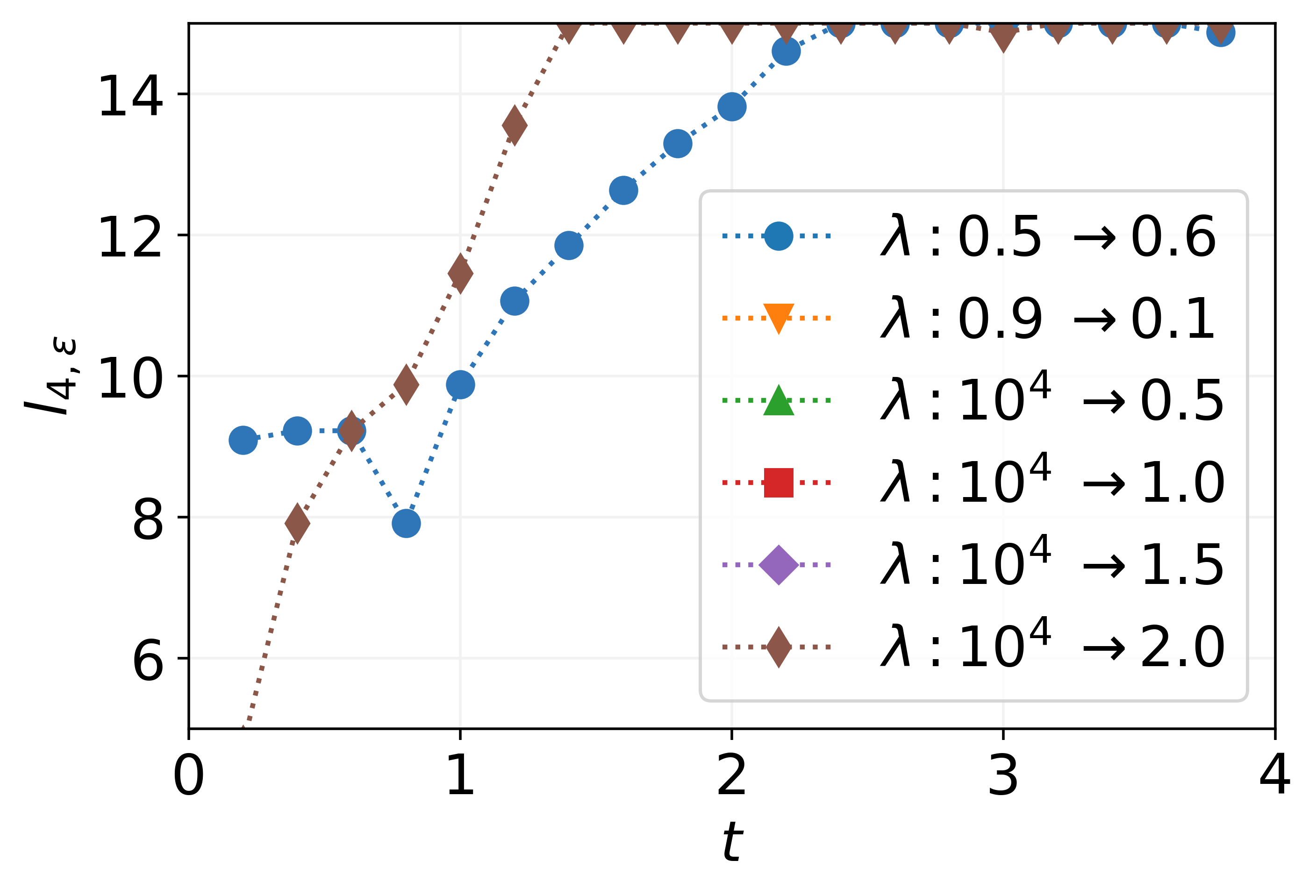}
  \hspace*{0.7cm}
  \end{subfigure}
  \caption{\raggedright{Evolution of the localization lengths $l_{2,\epsilon}$ and $l_{4,\epsilon}$ such that $\lvert \partial_L^2T_2\rvert\leq \epsilon$,
$\lvert \partial_L^2T_4\rvert\leq \epsilon$, for different quench protocols and time ranging from $0$ to $4$. We consider large quenches from $\lambda\gg 1$ and small quenches $\lambda'=\lambda+0.1$. Quenches to $\lambda=1.0$ are critical. The spreading is ballistic and defines a light-cone.}}
    \label{fig:length_evolution}
\end{figure}

Recently, there has been an intensive effort for the characterization of systems for which  SE can be computed efficiently~\cite{oliviero_magic-state_2022,haug_quantifying_2023,lami_quantum_2023,haug_stabilizer_2023}. In particular, for translationally invariant ground states of geometrically (gapped) local Hamiltonians~\cite{oliviero_magic-state_2022}, which can be well-described by Matrix Product States (MPSs), and in general for any MPS~\cite{haug_quantifying_2023}, there exists a constant $L_0$, the \textit{SE length}, such that for any $L>L_0$
\be
M_{2}(\rho_L)\simeq\alpha L+\beta
\label{additivemagic}
\ee
up to a small additive error $\epsilon\ll \alpha L+\beta$. In Eq.~\eqref{additivemagic} $\alpha,\beta,$ are constants that depend on the whole system state $\ket{\psi}$~\cite{oliviero_magic-state_2022} and therefore independent of the subsystem size $L$.  The linearity of the SE is a consequence of the finite correlation length of the state\cite{Hastings2006}. More precisely (see Section I of the Supplemental Material) the correction $\epsilon_L$ to the linear behavior in Eq.~\eqref{additivemagic} scales as $\epsilon_L=\gamma e^{-L/\xi}$, where $\gamma,\xi$ are constants depending on the finitely correlated state under consideration. Such behavior effectively defines the SE length being the constant $L_0$ such that $L_0=\xi\log\gamma/\epsilon$, for some tolerance $\epsilon\ll \alpha L_0+\beta$.

The existence of a (finite) SE length $L_0$ makes  SE easily computable for extended systems. To see this more concretely, consider a subsystem $X$ of size $L$. From Eq.~\eqref{additivemagic} we see that for $L>L_0+1$ and $\Delta \equiv M_{2}(\rho_{L_0+1})-M_{2}(\rho_{L_0})$ one has
\be
M_{2}(\rho_L)\simeq \Delta(L-L_0)+M_{2}(\rho_{L_0})\label{proportionalityeq}
\ee
%\beM_{2}(\rho_{L_1})\simeq \frac{L_2}{L_1}M_{2}(\rho_{L_2})\ee
which tells us that, once SE is measured for two subsystems of sizes $L_0$ and $L_0+1$ it can then be efficiently extrapolated, through Eq.~\eqref{proportionalityeq}, to a larger system sizes $L$. Note that it is crucial that $L>L_0$ to ensure the validity of Eq.~\eqref{additivemagic} and thus of Eq.~\eqref{proportionalityeq}.  
The SE length $L_0$ quantifies both  how non-stabilizerness is localized in the system  and the effort needed to compute SE. As an example, in the ground state $\ket{\psi_0(\lambda)}$ of the TFIM, the SE length  is $L_0=1$ for every $\lambda\gg 1$ and $\lambda\ll 1$~\cite{oliviero_magic-state_2022}.

%%%%%%%%%%%%%%%%%%%%%%%%%%%%%%%%%%%%%%%%%%%%%%%%%%%%%%%%%%%%%%%%%%%%%%%%%%%%%%%%%%%%%%%%%%%%%%%%%%%%%%%%%%%%%%%%%%%%%%%%%%%%%%%%%%%%%%%%%%%%%%%%%%%%%%%%%%%%%%%%%%%%%%%%%%%%%%%%%%%%%%%%%%%%%%%%%%%%%%%%%%%%%%%%%%%%%%%%%%%%%%%%%%%%%%%%%%%%%%%%%%%%%%%%%%%%%%%%%%%%%%%%%%%%%%%%%%%%%%%%%%%%%%%%%%%%%%%%%%%%%%%%%%%%%%%%%%%%%%%%%%%%%%%%%%%%%%%%%%%%%%%%%%%%%%%%%%%%%%%%%%%%%%%%%%%%%%%%%%%%%%%%%%%%%%%%%%%%%%%%%%%%%%%%%%%%%%%%%%%%%%%%%%%%%%%%%%%%%%%%%%%%%%%%%%%%%%%%%%%%%%%%%%%%%%%%%%%%%%%%%%%%%%%%%%%%%%%%%%%%%%%%%%%%%%%%%%%%%%%%%%%%%%%%%%%%%%%%%%%%%%%%%%%%%%%%%%%%%%%%%%%%%%%%%%%%%%%%%%%%%%%%%%%%%%%%%%%%%%%%%%%%%%%%%%%%%%%%%%%%%%%%%%%%%%%%%%%%%%%%%%%%%

\subsection{Nonstabilizerness delocalization}\label{subsec:SE_deloc}

We have seen that after a quantum quench, $M_2$ equilibrates after a time scaling linearly with the size of the system. This suggests that SE is spreading throughout the system. Such spreading should result in an increase in SE length. The main goal of this section is to show that the growth of the SE length $L_t$ is upper bounded by an effective light-cone.

We consider the case of states with finite correlation lengths. As it is well known, such states admit an efficient description by MPS\cite{orus_practical_2014}. Their time evolution under a local Hamiltonian $H(\lambda')$ results in a spreading of correlations\cite{nachtergaele_propagation_2006} and  increasing entropy of subsystems\cite{eisert2013entanglement}. Both effects are encoded in the bond dimensions $D$ for the MPS description, which increases at most as $D(t)\leq e^{A+ v t}$\cite{Alhambra2021tensor}, where $v$ is $O(1)$ in the system size. Using the fact that both the purity $\pur(\rho_L)$ and the stabilizer purity $W(\rho_L)$ can be written as a expectation value of a string of local observables on the replica state $\ket{(\psi\otimes \psi^{*})^{\otimes k}}$ for $k=1,2$ respectively~\cite{haug_quantifying_2023}, we can show that  
the SE length $L_t$ obeys the following bound
\be
L_t\le L_0+v_{s}t\,.
\label{mainresult}
\ee
We refer to Section I of the Supplemental Material for details. Here $v_s$ is a constant that plays the role of an effective velocity and is $O(1)$ in the system size. 
The above equation %is one of the main result of the paper: it 
shows that, under the evolution by a local Hamiltonian, SE delocalizes within an effective light-cone constrained by the finite range of interactions in the quench Hamiltonian $H(\lambda^{\prime})$. %Moreover, a direct relationship between SE and the bond dimension $D$ is established.

%%%%%%%%%%%%%%%%%%%%%%%%%%%%%%%%%%%%%%%%%%%%%%%%%%%%%%%%%%%%%%%%%%%%%%%%%%%%%%%%%%%%%%%%%%%%%%%%%%%%%%%%%%%%%%%%%%%%%%%%%%%%%%%%%%%%%%%%%%%%%%%%%%%%%%%%%%%
%%%%%%%%%%%%%%%%%%%%%%%%%%%%%%%%%%%%%%%%%%%%%%%%%%%%%%%%%%%%%%%%%%%%%%%%%%%%%%%%%%%%%%%%%%%%%%%%%%%%%%%%%%%%%%%%%%%%%%%%%%%%%%%%%%%%%%%%%%%%%%%%%%%%%%%%%%%
%%%%%%%%%%%%%%%%%%%%%%%%%%%%%%%%%%%%%%%%%%%%%%%%%%%%%%%%%%%%%%%%%%%%%%%%%%%%%%%%%%%%%%%%%%%%%%%%%%%%%%%%%%%%%%%%%%%%%%%%%%%%%%%%%%%%%%%%%%%%%%%%%%%%%%%%%%%
%%%%%%%%%%%%%%%%%%%%%%%%%%%%%%%%%%%%%%%%%%%%%%%%%%%%%%%%%%%%%%%%%%%%%%%%%%%%%%%%%%%%%%%%%%%%%%%%%%%%%%%%%%%%%%%%%%%%%%%%%%%%%%%%%%%%%%%%%%%%%%%%%%%%%%%%%%%%%%%%%%%%%%%%%%%%%%%%%%%%%%%%%%

\subsection{SE length growth in the TFIM}\label{subsec:SE_deloc_TFIM}

In this subsection, we compute in the concrete case of TFIM the speed $v_s$ for the SE length growth, that is, its delocalization.
%Let us now turn to analyze nonstabilizerness delocalization. 
The behavior of the SE length $L_t$ is dictated by the correction to the linear scaling in Eq.~\eqref{additivemagic}. Recall that Eq.~\eqref{additivemagic} holds up to an additive error $\epsilon_{L_t}$, where now $L_t$ is a function of time $L_t\equiv L_{0}(t)$. The speed at which the error $\epsilon_{L_t}$ increases determines the velocity $v_s$ of the spreading of $L_t$ for $H_{I}(\lambda)$ (see Eq.~\eqref{mainresult}). We thus numerically investigate the second derivative with respect to $L$ of $M_{2}(\lambda,t)\equiv M_{2}(\rho_L^t(\lambda))$, for $\rho_L^t(\lambda)=\tr_{N\setminus L}\st{\psi_t(\lambda)}$ that, from  Eq.~\eqref{eq:SE_def}, is a sum of two terms
\be
\partial_{L}^2M_{2}(\lambda,t)=\partial_L^2T_{4}(\lambda,t)-\partial_{L}^2T_2(\lambda,t)
\label{secondderivative}
\ee
where $T_{4}(\lambda,t)\equiv-\log W(\rho_L^t(\lambda))-L$, and $T_{2}(\lambda,t)\equiv S_{2}(\rho_L^t(\lambda))-L$.  Eq.~\eqref{secondderivative} tells us that the behavior of $\epsilon_{L_t}$ is, ultimately, determined by the fastest (in $t$) contribution between $\partial_L^2T_{2}(\lambda,t)$ and $\partial_L^2T_{4}(\lambda,t)$. Interestingly, the second term $\partial_L^2T_{2}(\lambda,t)$  encodes the sublinear correction for the area law of MPSs for the $2$-R\'enyi entropy of entanglement. In other words, the speed at which entanglement spread out in the system can be determined by only the behavior of $\partial_L^2T_{2}(\lambda,t)$. On the other hand, the corrections to the nonstabilizerness additivity in Eq.~\eqref{additivemagic} and their behavior after the quench are encoded in the evolution of both $\partial_L^2T_{4}(\lambda,t)$ and $\partial_L^2T_{2}(\lambda,t)$, which suggests a tight relationship between nonstabilizerness delocalization and entanglement growth. Indeed through theoretical arguments, in Section I of the Supplemental Material, we argue that nonstabilizerness could delocalize two times faster than entanglement. This consideration is found to be true by the numerical analysis below.

%%%%%%%%%%%%%%%%%%%%%%%%%%%%%%%%%%%%%%%%%%%%%%%%%%%%%%%%%%%%%%%%%%%%%%%%%%%%%%%%%%%%%%%%%%%%%%%%%%%%%%%%%%%%%%%%%%%%%%%%%%%%%%%%%%%%%%%%%%%%%%%%%%%%%%%%%%%%%%%%%%%%%%%%%%%%%%%%%%%%%%%%%%%%%%%%%%%%%%%%%%%%%%%%%%%%%%%%%%%%%%%%%%%%%%%%%%%%%%%%%%%%%%%%%%%%%%%%%%%%%%%%%%%%%%%%%%%%%%%%%%%%%%%%%%%%%%%%%%%%%%%%%%%%%%%%%%%%%%%%%%%%%%%%%%%%%%%%%%%%%%%%%%%%%%%%%%%%%%%%%%%%%%%%%%%%%%%%%%%%%%%%%%%%%%%%%%%%%%%%%%%%%%%%%%%%%%%%%%%%%%%%%%%%%%%%%%%%%%%%%%%%%%%%%%%%%%%%%%%%%%%%%%%%%%%%%%%%%%%%%%%%%%%%%%%%%%%%%%%%%%%%%%%%%%%%%%%%%%%%%%%%%%%%%%%%%%%%%%%

To extract the behavior of $L_t$ and thus of the velocity $v_s$, we fix an error tolerance $\epsilon$ and define $l_{2,\epsilon}(t)$ and $l_{4,\epsilon}(t)$ as solutions of the following inequalities $\lvert \partial_L^2T_2\rvert\leq \epsilon$,
$\lvert \partial_L^2T_4\rvert\leq \epsilon$ definitively in time. Then we extract $v_s=\max\{v_{T_2},v_{T_4}\}$ where $v_{T_k}\equiv \partial_{t}l_{k,\epsilon}$ for $k=2,4$. In Fig.~\ref{fig:ddT} we depict the behavior of $|\partial_L^2T_2(\lambda,t)|$ and $|\partial_L^2T_4(\lambda,t)|$ for the quench protocols in which the definitive behavior is captured with the computational resources at our disposal, see also see Section II of the Supplemental Material. As expected, we observe that $|\partial_L^2T_2(\lambda,t)|$ and $|\partial_L^2T_4(\lambda,t)|$ decay exponentially below the dotted line representing the chosen error tolerance $\epsilon$. The associated lengths $l_{2,\epsilon}(t)$ and $l_{4,\epsilon}(t)$ as a function of $t$ are shown in Fig.~\ref{fig:length_evolution} for $t=0,\ldots,4$: after an initial transient, both $l_{2,\epsilon}(t)$ and $l_{4,\epsilon}(t)$ grow ballistically. The associated velocities are: $v_{T2}\approx 2.5$ for $\lambda'=0.5$ and $v_{T2}\approx 5$ for $\lambda'\in\{1,1.5,2\}$, independently of the initial state; while $v_{T4}\approx 5$ for $\lambda'=0.5$ and $v_{T4}\approx 10$ for $\lambda'=2.0$. We conclude that, being $v_{T_4}>v_{T_2}$, $v_s\approx v_{T_{4}}$ and thus --  compatibly with our theoretical considerations -- the SE delocalizes two times faster than entanglement entropy. 

Finally, we are going to show that the velocity $v_s$ is proportional to the Lieb-Robinson speed $v_{LR}$. The Lieb-Robinson velocity associated with the Ising Hamiltonian can be reconstructed from the revivals after the quench. Revivals are brief detachments from the average value observables, whose magnitude decays in time as the equilibration process nears completion. During these detachments, the system state gets briefly closer to the initial state. Therefore, revivals can be detected by looking at the Loschmidt echo (LE), that is, the squared fidelity between the evolved state and the initial state. The revival times $T_\text{rev}$ are proportional to the system size $N$ and are related to maximal group velocity in integrable systems, and therefore to the Lieb-Robinson speed in generic local systems, as $T_\text{rev}\approx N/(2 v_\text{LR})$~\cite{happola_universality_2012}. The LE can be efficiently calculated for integrable spin chains in the fermionic representation described in Ref.~\cite{lieb_two_1961}. In Appendix~\ref{app:lr_speed} we show that, independently of the initial state, Lieb-Robinson speed is $v_\text{LR}=2$ for the paramagnetic quench $\lambda'>1$, and $v_\text{LR}=2 \lambda'$ for the ferromagnetic quench $\lambda'<1$. This result suggests a proportionality relation between $v_s$ and $v_\text{LR}$.

\section{Conclusions}\label{sec:conclusions}

In this work, we studied for the first time the behavior away from equilibrium of the non-stabilizer properties of a quantum many-body system after a sudden quench. The system studied is the integrable quantum Ising chain, which allowed for a thoroughly analytical treatment, and non-stabilizerness is computed through the Stabilizer R\'enyi Entropy $M_2$. 
Two main results are found: (i)  $M_2$ increases and finally equilibrates in a time proportional to the system size, and (ii) one can define a stabilizer entropy {\em length} $L_t$ that describes the SE localization~\cite{oliviero_magic-state_2022}. Such length increases linearly in time showing that $M_2$ spreads ballistically through the system until complete delocalization.  

In perspective, this work calls for several questions. In\cite{leone_quantum_2021,oliviero_transitions_2021, oliviero_black_2022, leone_learning_2022, leone_retrieving_2022, leone_phase_2023} it has been shown that the onset of quantum chaotic behavior in quantum circuits corresponds to a value of $M_2>N/2$, and full-fledged quantum chaos is attained near-maximal values for SE. Of course, chaos in quantum circuits is not the same than chaos in Hamiltonian systems, being defined there as the onset of universal entanglement features or universal behavior of the OTOCs\cite{leone_quantum_2021}. However, our results show that the equilibrium value of SE for the integrable quantum quench is below the quantum chaotic threshold of quantum circuits. This fact raises the question whether the equilibrium SE is a tell-tale of the onset of quantum chaos in Hamiltonians system. We indeed speculate 
that a non-integrable system will equilibrate to a larger value for $M_2$. It would be intriguing if the increase in equilibrium SE compared to the integrable case would depend on the strength of the interability breaking term. 

Second, preliminary numerical analysis suggests that subtle features of SE dynamics may be erased by the operation of partial trace. This is akin to the problem of presence of thermal fluctuations when evaulating entanglement. A possible strategy would be to localize SE in a subsystem by measurements in a Clifford basis (e.g., the computational basis) and evaluate the local residual SE in the pure state. This would entail to average over all the possible Clifford measurements. 

Third, the onset of quantum chaos depends on the interplay between both entanglement and SE, \cite{leone_quantum_2021,leone_retrieving_2022,oliviero_transitions_2021,dowling_scrambling_2023,leone_isospectral_2021,oliviero_random_2021,leone_learning_2022,oliviero_black_2022, dowling_quantum_2023,omanakuttan_scrambling_2023,goto_probing_2022,true_transitions_2022,kim_quantum_2023} and it is still an open question to what extent they are sufficient and/or necessary. Recently\cite{tirrito_quantifying_2023}, it has been shown that the flatness of the entanglement spectrum of a subsystem is a good probe for SE. Since it is also a probe for entanglement, it is tempting to study the dynamics of flatness to probe the onset of quantum chaos.

Finally, finite size scaling in $L$ for larger sizes would allow for a reliable analysis of the temporal fluctuations. To this end, we plan to employ Monte-Carlo methods to sample SE efficiently. Finally, we remark on the fascinating relationship between SE and the bond dimension $D$. The study of the interplay of SE with the efficiency of  tensor network methods is thus potentially of great importance for the issue of simulating quantum many-body systems on a classical computer.

\section*{Acknowledgments}

The authors acknowledge important discussions with R. Fazio, P. Lucignano and J. Odavic. AH was  supported by the PNRR MUR project PE0000023-NQSTI and PNRR MUR project CN $00000013$-ICSC.  L.L., S.O. acknowledge support from NSF award number 2014000. D.R. acknowledges support from the EU project EURyQA.

%%APPENDICES---------------------

\appendix

\section{Proof of Eq.~(6)}\label{App:proofofthemainresults}

First of all, let us define the functions $T_{4}(\lambda,t)\equiv-\log W(\rho_L^t(\lambda))-L$, and $T_{2}(\lambda,t)\equiv S_{2}(\rho_L^t(\lambda))-L$. In terms of $T_2$ and $T_4$, the SE can be written as
\begin{align}
M_{2}(\lambda,t)=T_{4}(\lambda,t)-T_2(\lambda,t).
\end{align}

As shown in \cite{haug_quantifying_2023}, $2^{T_2}$ and $2^{T_4}$ are expectation values of strings of $L$ connected observables on replica states. More precisely, define $\ket{\psi^{(k)}}\equiv\ket{(\psi\otimes\psi^*)^{\otimes k}}$ for $k=1,2$. Define the local observable $A_{i}^{(k)}=\bbbone_{i}^{\otimes 2k}+\sum_{\alpha=1}^{3}(\sigma_{i}^{\alpha}\otimes\sigma_{i}^{\alpha*})^{\otimes k}$ for $k=1,2$ and $\sigma_{i}^{\alpha}$ for $\alpha=x,y,z$ being single qubit Pauli matrices. Then $2^{T_{2k}}$ for $k=1,2$ can be recast as
\begin{align}
2^{T_{2k}(\lambda,t)}&=\bra{\psi^{(k)}}A^{(k)}_1\otimes\dots\otimes A^{(k)}_L\otimes\bbbone_{N\setminus L}\ket{\psi^{(k)}}.
\end{align}

To explain the behavior of the SE under time evolution and its relation to the locality, we want to understand how strings of observables behave in a state with finite correlation length, and how they change while, after a quantum quench, correlations spread out in the system.

We start by looking at the initial translationally invariant state $\ket{\psi(0)}$. Having a finite correlation length, this can be represented as an MPS with polynomial bond dimension $D$. Similarly, $\ket{\psi^{(k)}}$ for $k=1,2$ is a MPS state with bond dimension $D_{k}=D^{2k}$.  Let $\tau_{\psi^{(k)}}$ be the transfer matrix of the MPS and $\tau_{\psi^{(k)}\!,A}$ be the transfer matrix of the MPS contracted with the local operator $A$. The expectation value of a string of local operators $A_i^{(k)}$ can be written as:
\begin{align}
&\bra{\psi^{(k)}}A^{(k)}_1\otimes\dots\otimes A^{(k)}_L\otimes\bbbone_{N\setminus L}\ket{\psi^{(k)}}\nonumber\\
&=\lim_{N\rightarrow\infty}\tr\left(\tau_{\psi^{(k)}}^{N-L}\tau_{\psi^{(k)}\!,A}^{L}\right)
\end{align}

Now, we call $\ket{R_{\psi^{(k)}}^{(i)}}$ and $\ket{L_{\psi^{(k)}}^{(i)}}$ the right and left eigenvectors of $\tau_{\psi^{(k)}}$ and $\lambda_{\psi^{(k)}}^{(i)}$ the associated eigenvalues. In this way, the transfer matrix can be written as
\begin{align}
    \tau_{\psi^{(k)}}=\sum_{i=1}^{D_k^2}\lambda_{\psi^{(k)}}^{(i)}\ket{R_{\psi^{(k)}}^{(i)}}\bra{L_{\psi^{(k)}}^{(i)}},
\end{align}
where $\lvert\lambda_{\psi^{(k)}}^{(i)}\rvert>\lvert\lambda_{\psi^{(k)}}^{(i+1)}\rvert$, and $\lambda_{\psi^{(k)}}^{(1)}=1$ for the normalization condition. Analogously, we can write
\begin{align}
    \tau_{\psi^{(k)}\!,A}=\sum_{i=1}^{D_k^2}\lambda_{\psi^{(k)}\!,A}^{(i)}\ket{R_{\psi^{(k)}\!,A}^{(i)}}\bra{L_{\psi^{(k)}\!,A}^{(i)}},
\end{align}
with $\lvert\lambda_{\psi^{(k)}\!,A}^{(i)}\rvert>\lvert\lambda_{\psi^{(k)}\!,A}^{(i+1)}\rvert$. The eventual degeneration of eigenvalues does not affect the general behavior of this proof. Hence, a string expectation value can be written as
\begin{align}
 &\bra{\psi^{(k)}}A^{(k)}_1\otimes\dots\otimes A^{(k)}_L\otimes\bbbone_{N\setminus L}\ket{\psi^{(k)}}\nonumber\\
 &=\sum_{i=1}^{D_k^2}\left(\lambda_{\psi^{(k)}\!,A}^{(i)}\right)^L\bra{L_{\psi^{(k)}}^{(i)}} R_{\psi^{(k)}\!,A}^{(i)}\rangle\bra{L_{\psi^{(k)}\!,A}^{(i)}}R_{\psi^{(k)}}^{(i)}\rangle.
\end{align}

At this point, we define
\begin{align}
    2^{m(\psi^{(k)})}&:=\lambda_{\psi^{(k)}\!,A}^{(1)}\\
    2^{q(\psi^{(k)})}&:= \bra{L_{\psi^{(k)}}^{(1)}} R_{\psi^{(k)}\!,A}^{(1)}\rangle\bra{L_{\psi^{(k)}\!,A}^{(1)}}R_{\psi^{(k)}}^{(1)}\rangle\\
    e^{-1/\xi(\psi^{(k)})}&:= \left\lvert\frac{\lambda_{\psi^{(k)}\!,A}^{(2)}}{\lambda_{\psi^{(k)}\!,A}^{(1)}}\right\rvert\\
    F(\psi^{(k)})&:= \left\lvert\frac{\bra{L_{\psi^{(k)}}^{(2)}} R_{\psi^{(k)}\!,A}^{(2)}\rangle\bra{L_{\psi^{(k)}\!,A}^{(2)}}R_{\psi^{(k)}}^{(2)}\rangle}{\bra{L_{\psi^{(k)}}^{(1)}} R_{\psi^{(k)}\!,A}^{(1)}\rangle\bra{L_{\psi^{(k)}\!,A}^{(1)}}R_{\psi^{(k)}}^{(1)}\rangle}\right\rvert.
\end{align}
Since $\left\lvert\lambda_{\psi^{(k)}\!,A}^{(i)}\right\rvert^L\gg\left\lvert\lambda_{\psi^{(k)}\!,A}^{(i+1)}\right\rvert^L$, for sufficiently large $L$ we obtain
\begin{align}\label{App:eqbond}
    &\bra{\psi^{(k)}}A^{(k)}_1\otimes\dots\otimes A^{(k)}_L\otimes\bbbone_{N\setminus L}\ket{\psi^{(k)}}\nonumber\\
    &=2^{m(\psi^{(k)})L+q(\psi^{(k)})}(1+\varepsilon^{(k)})
\end{align}
where 
\begin{align}
    \lvert\varepsilon^{(k)}\rvert\leq D_k^2 e^{-L/\xi(\psi^{(k)})} F(\psi^{(k)})\ll1.
\end{align}
It should be noted that the derivation of this bound assumes that $F(\psi^{(k)})$ is finite. In the event that this condition is not met, an equivalent derivation can be obtained by considering the next eigenvalue $\lambda_{\psi^{(k)}\!,A}^{(2)}$. In this case, $F(\psi_k):= \left\lvert\bra{L_{\psi^{(k)}}^{(3)}} R_{\psi^{(k)}\!,A}^{(3)}\rangle\bra{L_{\psi^{(k)}\!,A}^{(3)}}R_{\psi^{(k)}}^{(3)}\rangle\right\rvert/$ $\left\lvert \bra{L_{\psi^{(k)}}^{(2)}} R_{\psi^{(k)}\!,A}^{(2)}\rangle\bra{L_{\psi^{(k)}\!,A}^{(2)}}R_{\psi^{(k)}}^{(2)}\rangle\right\rvert$, and the linear approximation is preserved with the same exponential correction.

With this formalism, also the effect of locality on time evolution can be easily addressed. When an MPS evolves with a local Hamiltonian $H(\lambda')$, the spreading of correlations\cite{nachtergaele_propagation_2006} and the increasing entropy of subsystems\cite{eisert2013entanglement} are encoded in the bond dimensions, which increases at most as $D(t)\leq e^{A+ v t}$\cite{Alhambra2021tensor}, where $v$ is $O(1)$ in the system size. We can then generalize the result obtained in Eq.~\eqref{App:eqbond} in the following way
\begin{align}
    &\bra{\psi^{(k)}(t)}A^{(k)}_1\otimes\dots\otimes A^{(k)}_L\otimes\bbbone_{N\setminus L}\ket{\psi^{(k)}}\nonumber\\
    &=2^{m(\psi^{(k)}(t))L+q(\psi^{(k)(t)})}(1+\varepsilon^{(k)})
\end{align}
where we defined $\ket{\psi^{(k)}(t)}\equiv[U_{t}\otimes U_{t}^{*}]^{\otimes k}\ket{\psi^{(k)}}$ with $U_{t}=e^{-iH(\lambda')t}$. And consequently $\varepsilon^{(k)}$ is now bounded for sufficiently large $L$ as:
\begin{align}
    \lvert\varepsilon^{(k)}\rvert\leq e^{4kA+ 4kv t} e^{-L/\xi(\psi^{(k)}(t))} F(\psi^{(k)}(t))\ll1.
\end{align}
where we used that $D_k=D^{2k}$ and consequently that $D_k(t)=D(t)^{2k}\leq e^{2kA+2kvt}$.
At this point we define $\varepsilon_0^{(k)}=e^{4kA}\max_tF(\psi(t))$ and $\chi^{(k)}=\max_t\xi(\psi^{(k)}(t))$. We obtain that
\begin{align}
    \lvert\varepsilon^{(k)}\rvert\leq \varepsilon_0^{(k)} e^{(4kv\chi^{(k)} t-L)/\chi^{(k)}}.
\end{align}

With this result, let us focus on the behavior of the SE $M_2=T_4-T_2$. Since $2^{T_2}$ and $2^{T_4}$ are equal to
\begin{align}\label{eq:td_bonds}
    T_2&=m(\psi^{(1)}(t))L+q(\psi^{(1)}(t))+\varepsilon^{(1)}\nonumber\\
    T_4&=m(\psi^{(2)}(t))L+q(\psi^{(2)}(t))+\varepsilon^{(2)}\nonumber\\
    \lvert\varepsilon^{(1)}\rvert&\leq \varepsilon_0^{(1)} e^{(4v\chi^{(1)} t-L)/\chi^{(1)}}\nonumber\\
    \lvert\varepsilon^{(2)}\rvert&\leq \varepsilon^{(2)}_0 e^{(8v\chi^{(2)} t-L)/\chi^{(2)}}.
\end{align}

It follows that
\begin{align}
    M_2(\rho_L)&=[m(\psi^{(2)}(t))-m(\psi^{(1)}(t))]L\nonumber\\
    &+[q(\psi^{(2)}(t))-q(\psi^{(1)}(t))]+\varepsilon_M\nonumber\\
    \lvert\varepsilon_M\rvert &\leq 2\max\{\varepsilon^{(1)},\varepsilon^{(2)}\}.
\end{align}
In a nutshell, during time evolution corrections to the linearity of nonstabilizerness are exponentially small out from a light-cone. The speed at which the exponential bound increases determines the light-cone in which the SE spreads out. To figure this out, we fix an error tolerance $\epsilon$ and we find the values $L_{2,\epsilon}$ and $L_{4,\epsilon}$ such that $\lvert \varepsilon^{(1)}\rvert\leq \epsilon$ and
$\lvert \varepsilon^{(2)}\rvert\leq \epsilon$ definitively as a consequence of the previous bound. The SE length $L_t$ is $\max\{L_{2,\epsilon},L_{4,\epsilon}\}$.

Taking into account Eqs.(\ref{eq:td_bonds}) $L_{2,\epsilon}$ and $L_{2,\epsilon}$ increases at most as:
\begin{align}
L_{2,\epsilon}&=\xi(\psi^{(1)}(t))\log(\varepsilon_0^{(1)}/\epsilon)+4v\chi^{(1)} t\nonumber\\
L_{4,\epsilon}&=\xi(\psi^{(2)}(t))\log(\varepsilon^{(2)}_0/\epsilon)+8v\chi^{(2)} t
\end{align}
Therefore, we can conclude that 
\be
L_t\le L_0+v_{s}t\,,
\ee
where $v_{s}=\max\{4v\chi^{(1)},8v\chi^{(2)} \}$  is $O(1)$ in the system size. Remarkably, when $\chi^{(1)}$ is comparable with $\chi^{(2)}$, the error on $T_4$ spreads two times faster than the error on $T_2$.

\section{Additional numerical data}\label{App:additionalnumericaldata}

\begin{figure*}
  \begin{subfigure}[b]{0.25\textwidth}
    \includegraphics[height=65mm]{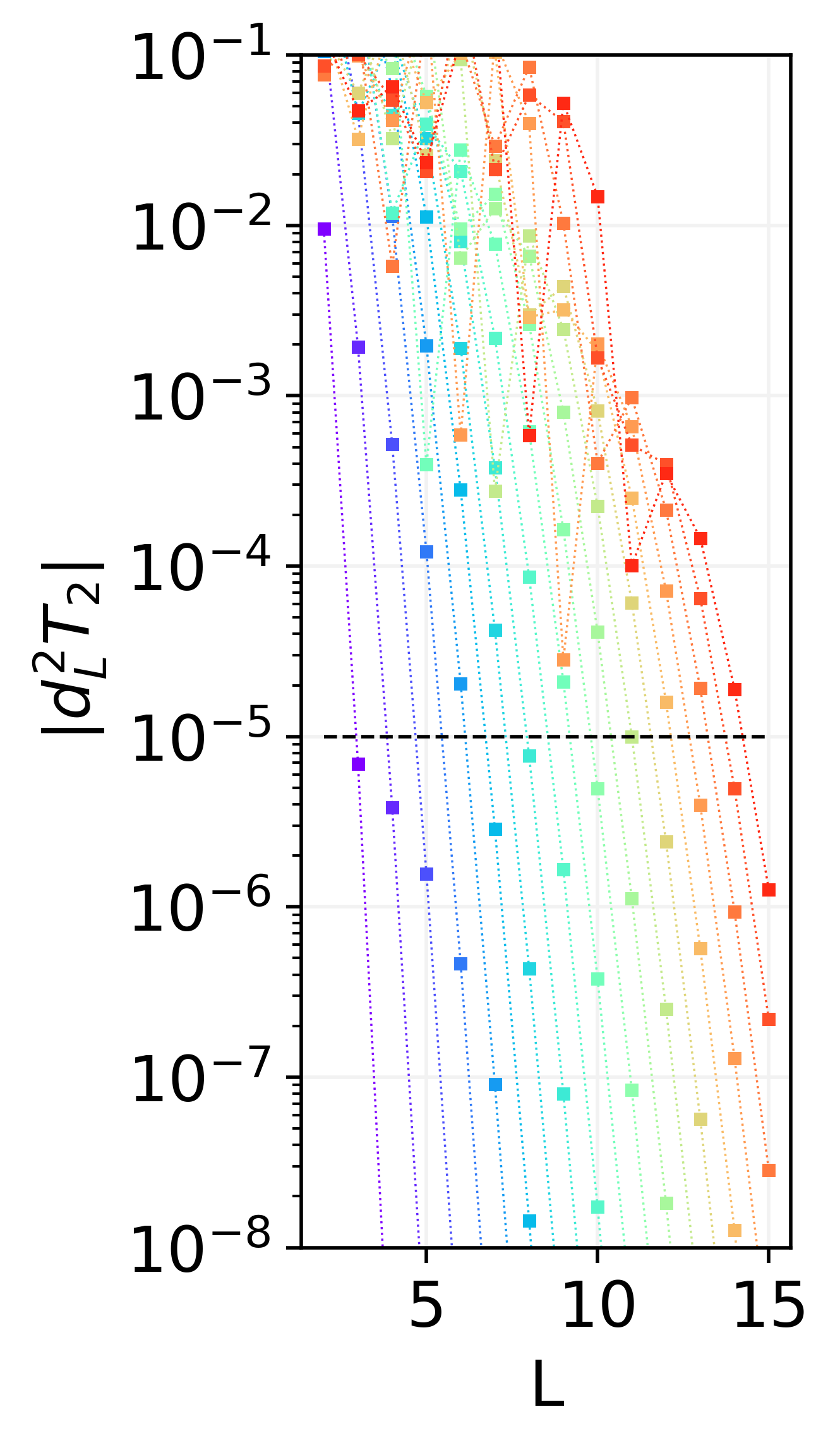}
    \caption{\kern-4em}
  \end{subfigure}\hfill
  \begin{subfigure}[b]{0.25\textwidth}
    \includegraphics[height=65mm]{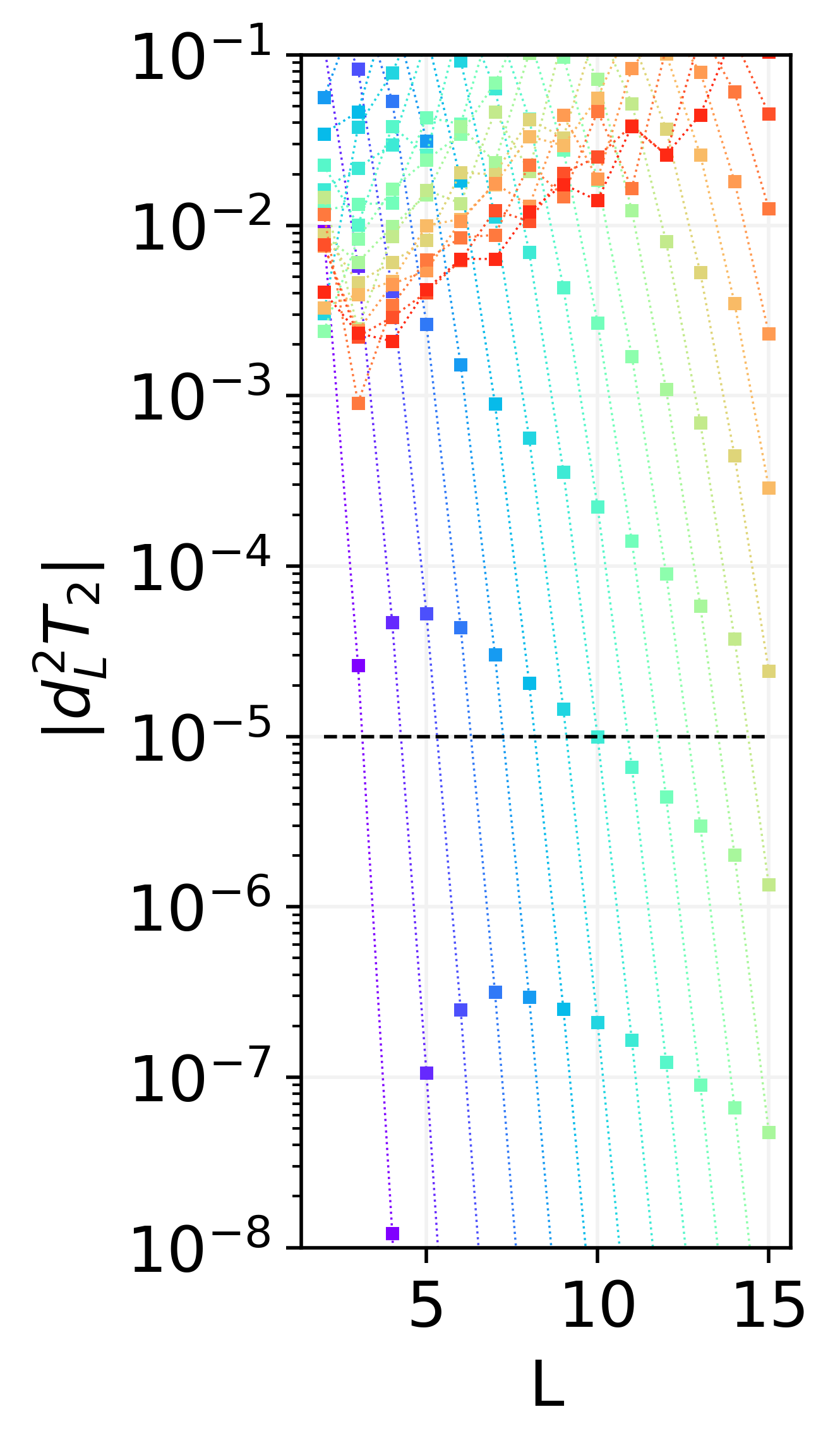}
    \caption{\kern-4em}
  \end{subfigure}\hfill
  \begin{subfigure}[b]{0.25\textwidth}
    \includegraphics[height=65mm]{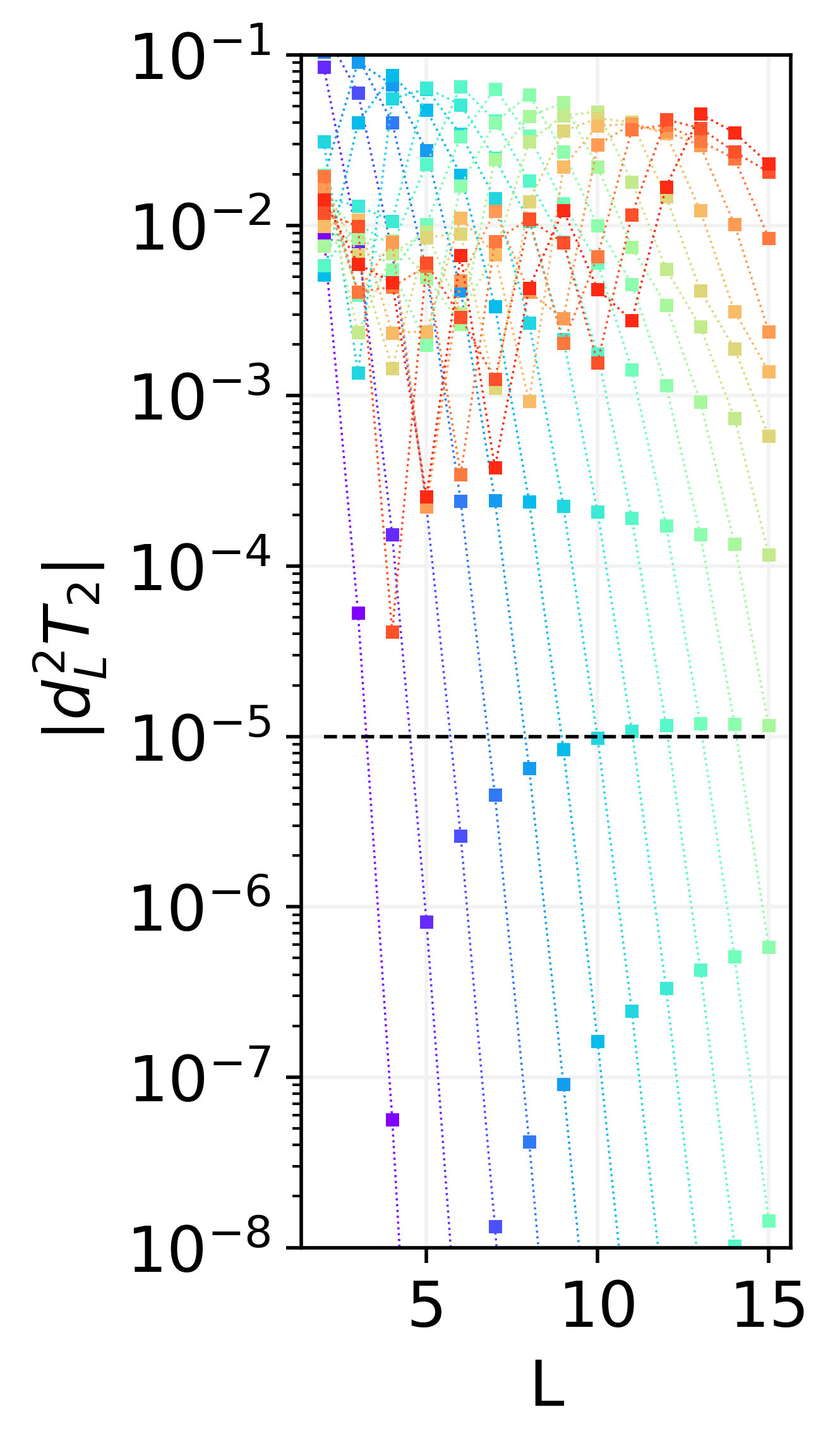}
    \caption{\kern-4em}
  \end{subfigure}\hfill
  \begin{subfigure}[b]{0.25\textwidth}
    \includegraphics[height=65mm]{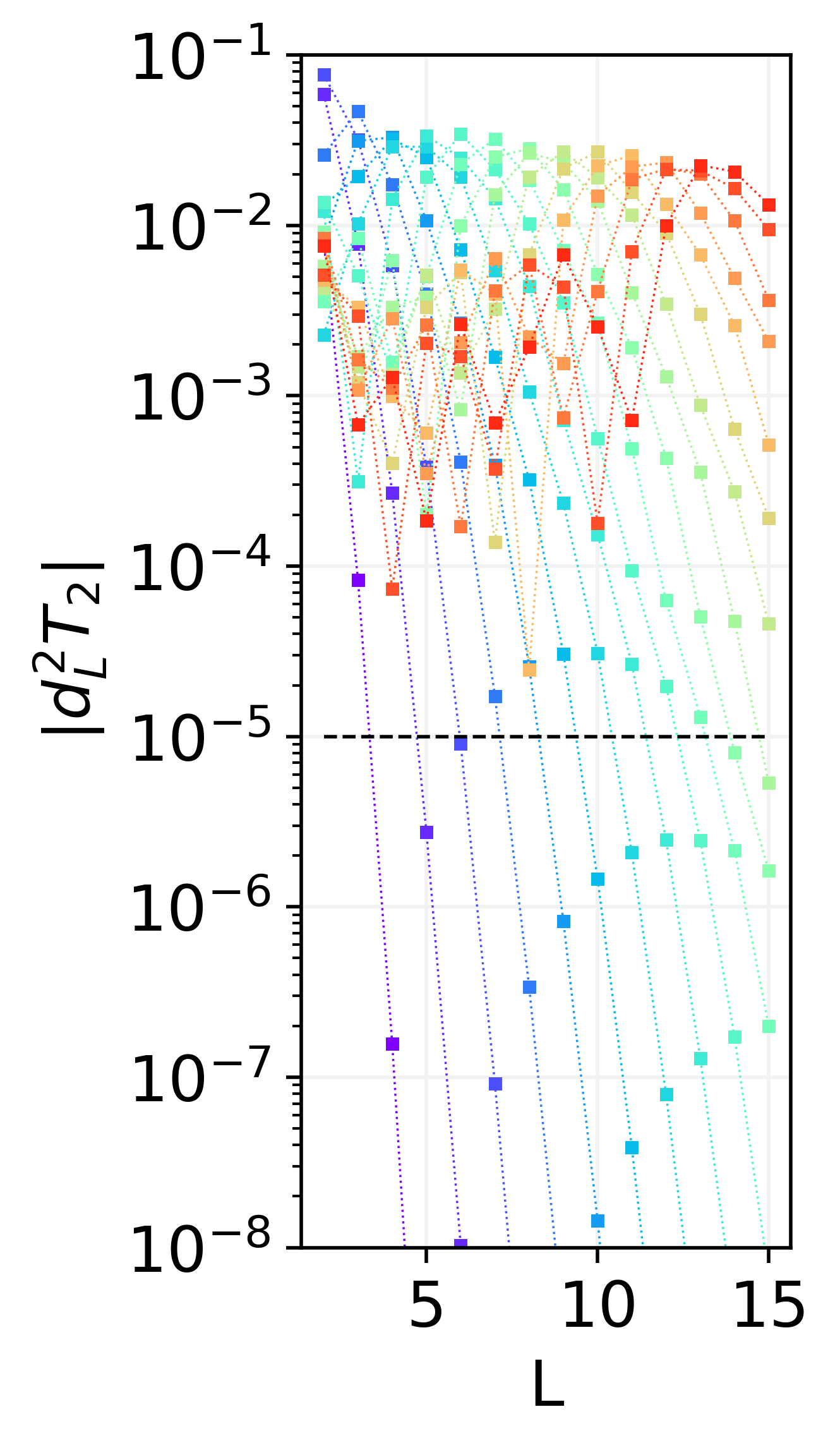}
    \caption{\kern-4em}
  \end{subfigure}
  \caption{\raggedright{Second space derivative of $T_2(\rho_L)$ as a function of the subsystem size $L$ after a quench from $\lambda=10^4$ to different values of $\lambda'$. The subsystem size goes from $2$ to $15$. Different lines represent different times in the range from $0$ (purple line below), to $4$ (red line above). The horizontal black line represents the error tolerance such that functions above the line are definitively exponential. In Panel (a) $\lambda'= 0.5$; in Panel (b) $\lambda'= 1.0$ (critical quench); in Panel (c) $\lambda'= 1.5$; in Panel (d) $\lambda'= 2.0$.}}
    \label{fig:allUpddT2}
\end{figure*}

\begin{figure*}
  \begin{subfigure}[b]{0.3\textwidth}
    \includegraphics[height=65mm]{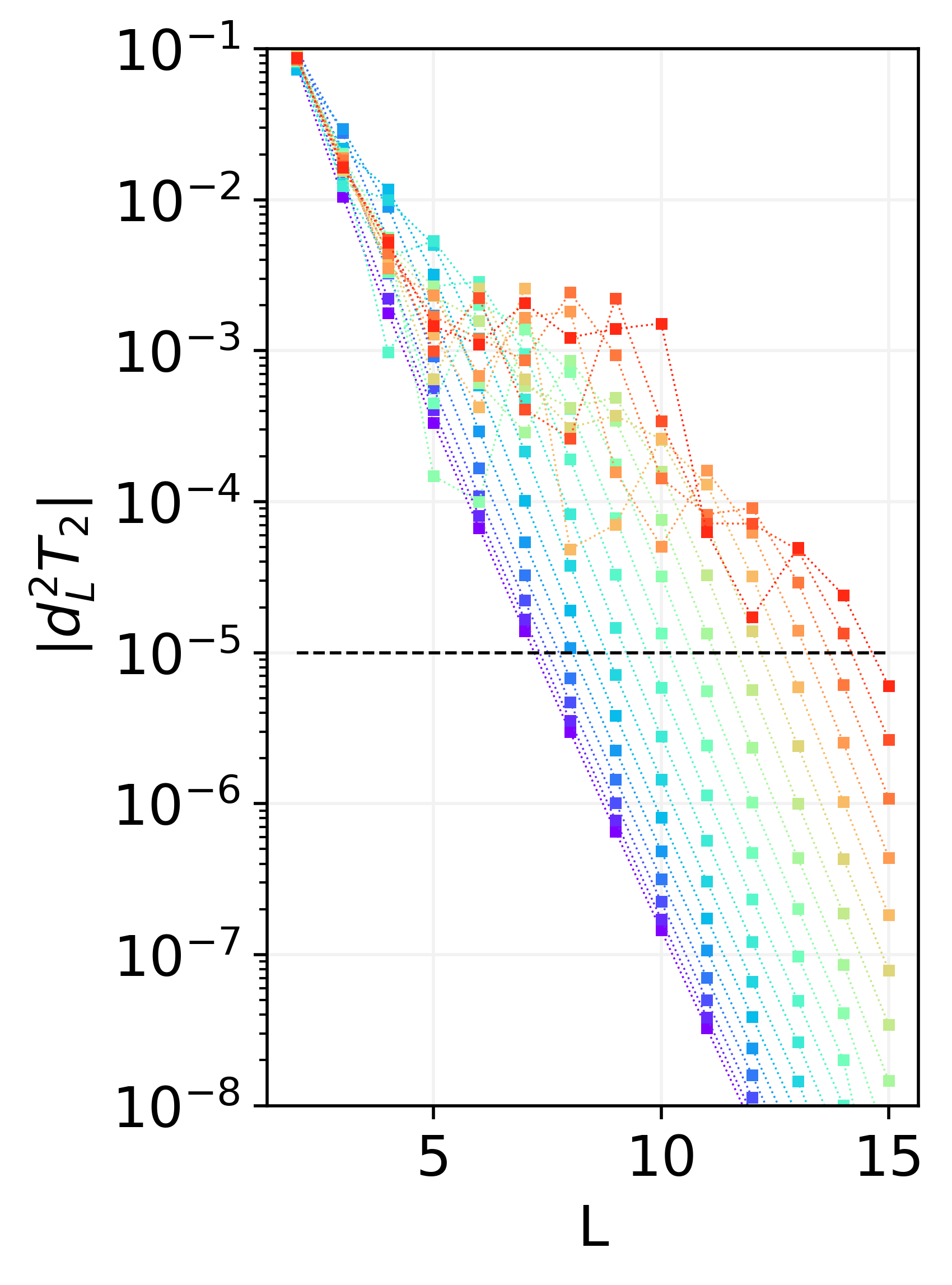}
    \caption{\kern-4em}
  \end{subfigure}\hfill
  \begin{subfigure}[b]{0.3\textwidth}
    \includegraphics[height=65mm]{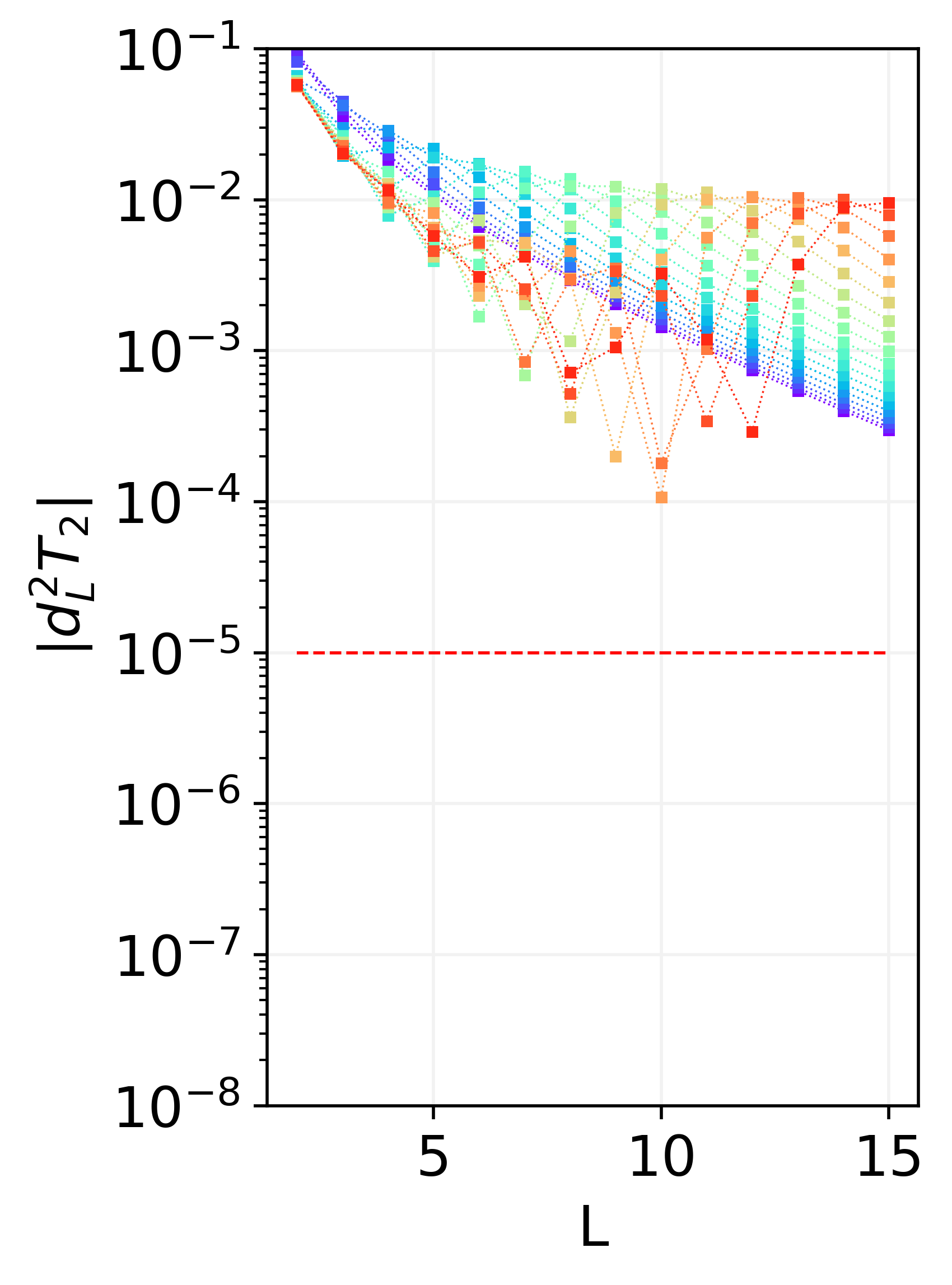}
    \caption{\kern-4em}
  \end{subfigure}\hfill
  \begin{subfigure}[b]{0.3\textwidth}
    \includegraphics[height=65mm]{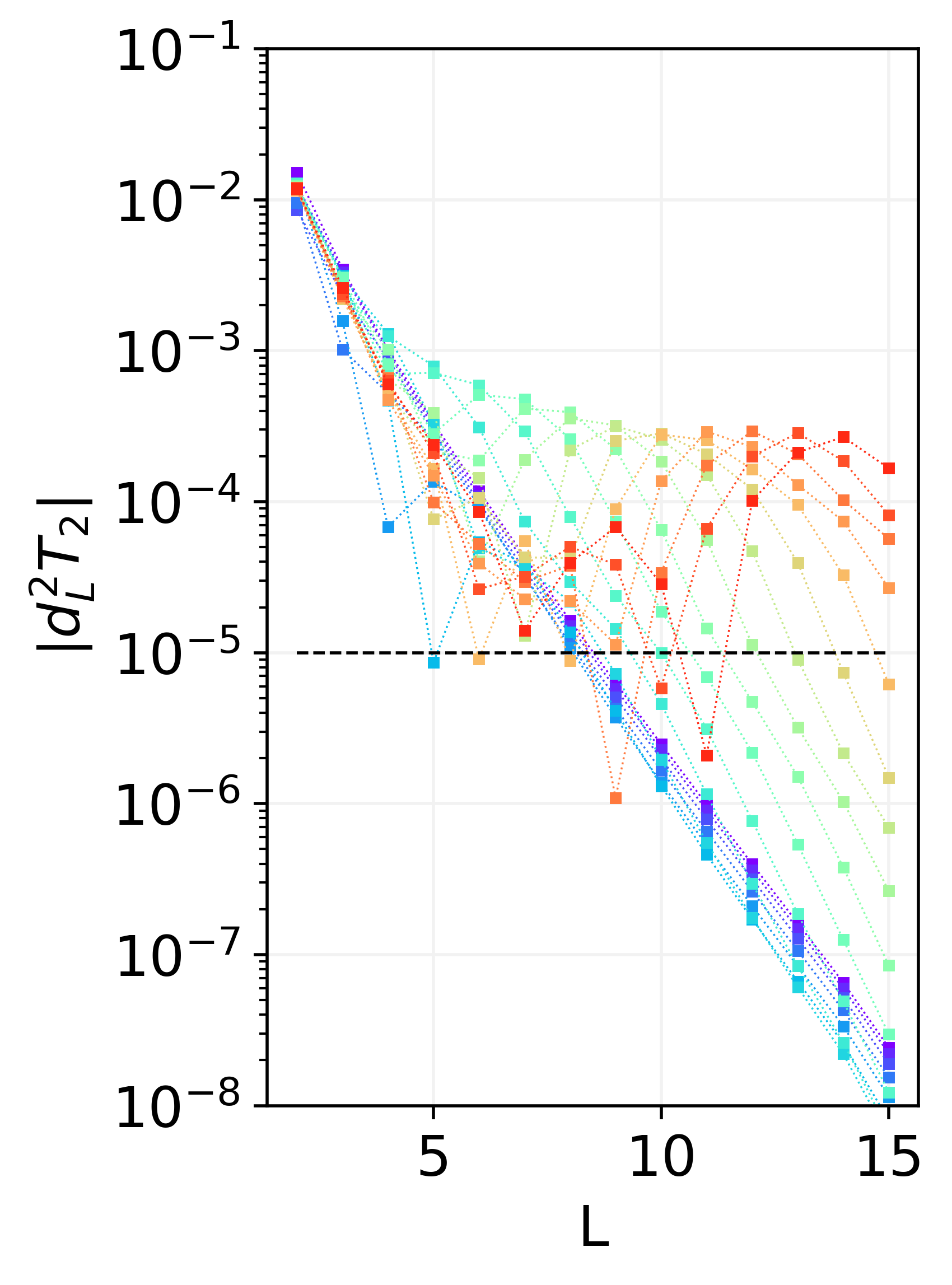}
    \caption{\kern-4em}
  \end{subfigure}
  \caption{\raggedright{Second space derivative of $T_2(\rho_L)$ as a function of the subsystem size $L$ after a quench from $\lambda$ to different values of $\lambda'=\lambda+0.1$. The subsystem size goes from $2$ to $15$. Different lines represent different times in the range from $0$ (purple line below), to $4$ (red line above). The horizontal black line represents the error tolerance such that functions above the line are definitively exponential. In Panel (a) $\lambda= 0.5$; in Panel (b) $\lambda= 0.9$ (critical quench); in Panel (c) $\lambda= 1.5$.}}   \label{fig:smallddT2}
\end{figure*}

\begin{figure*}
  \begin{subfigure}[b]{0.25\textwidth}
    \includegraphics[height=65mm]{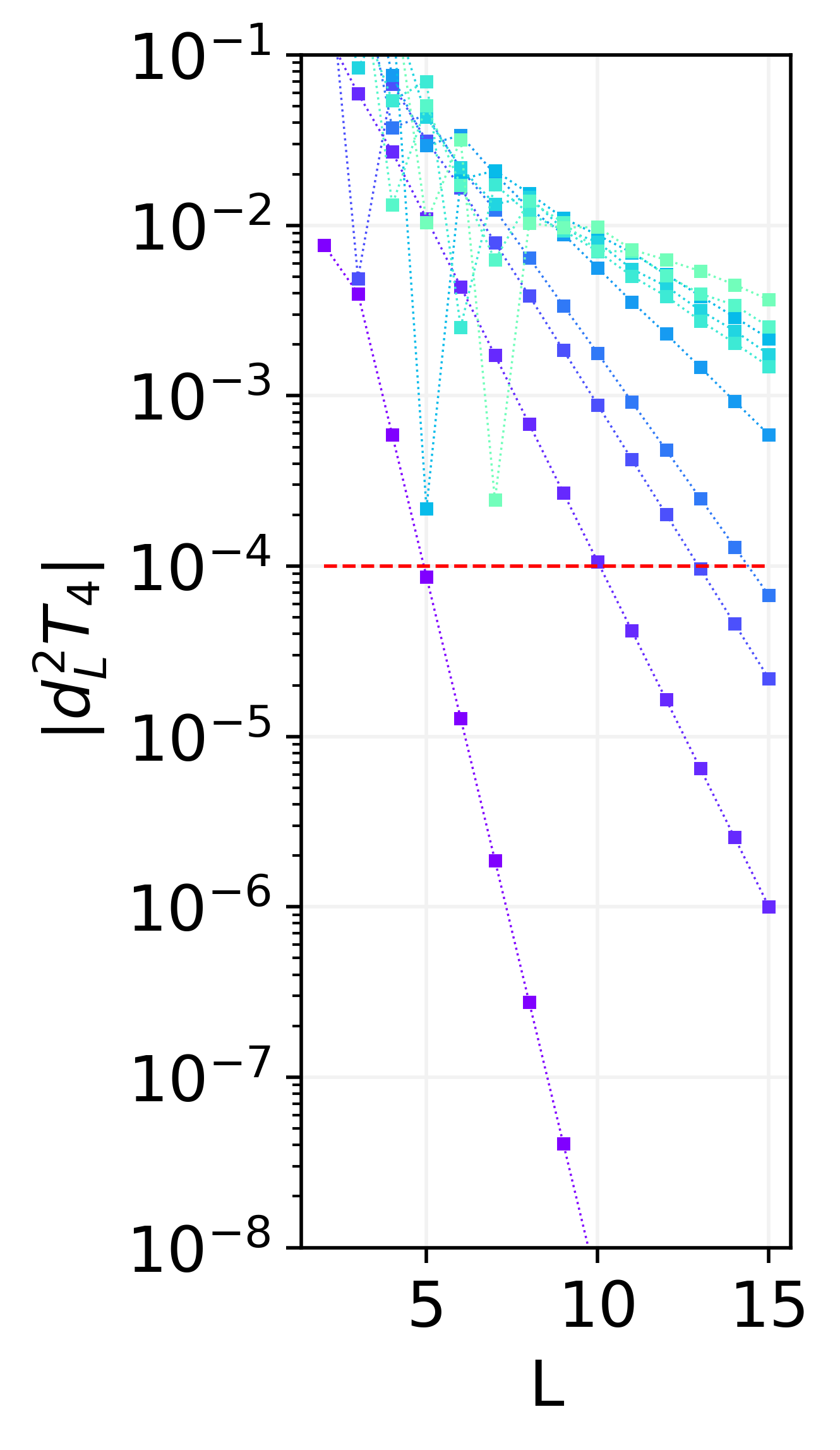}
    \caption{\kern-4em}
  \end{subfigure}\hfill
  \begin{subfigure}[b]{0.25\textwidth}
    \includegraphics[height=65mm]{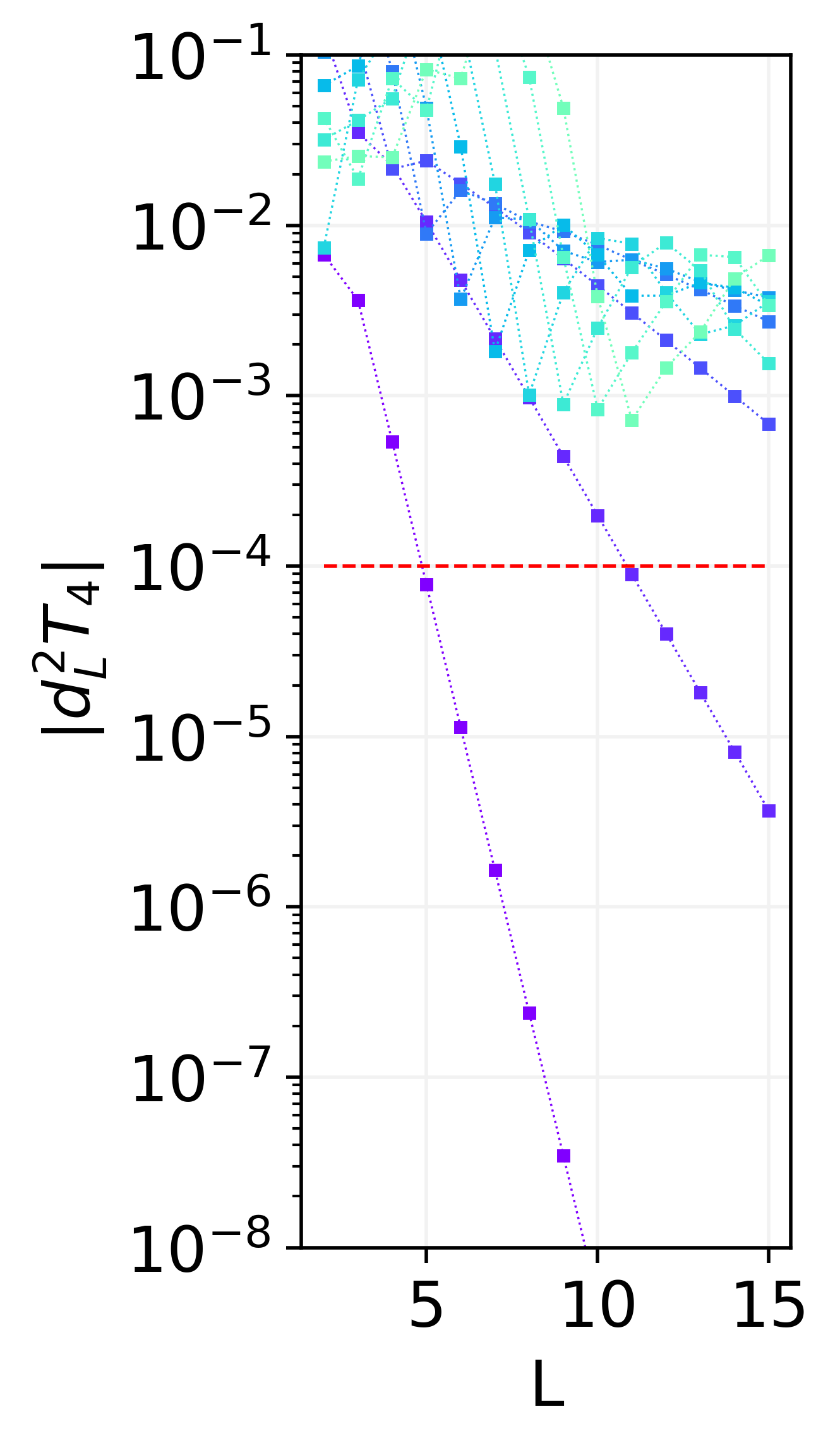}
    \caption{\kern-4em}
  \end{subfigure}\hfill
  \begin{subfigure}[b]{0.25\textwidth}
    \includegraphics[height=65mm]{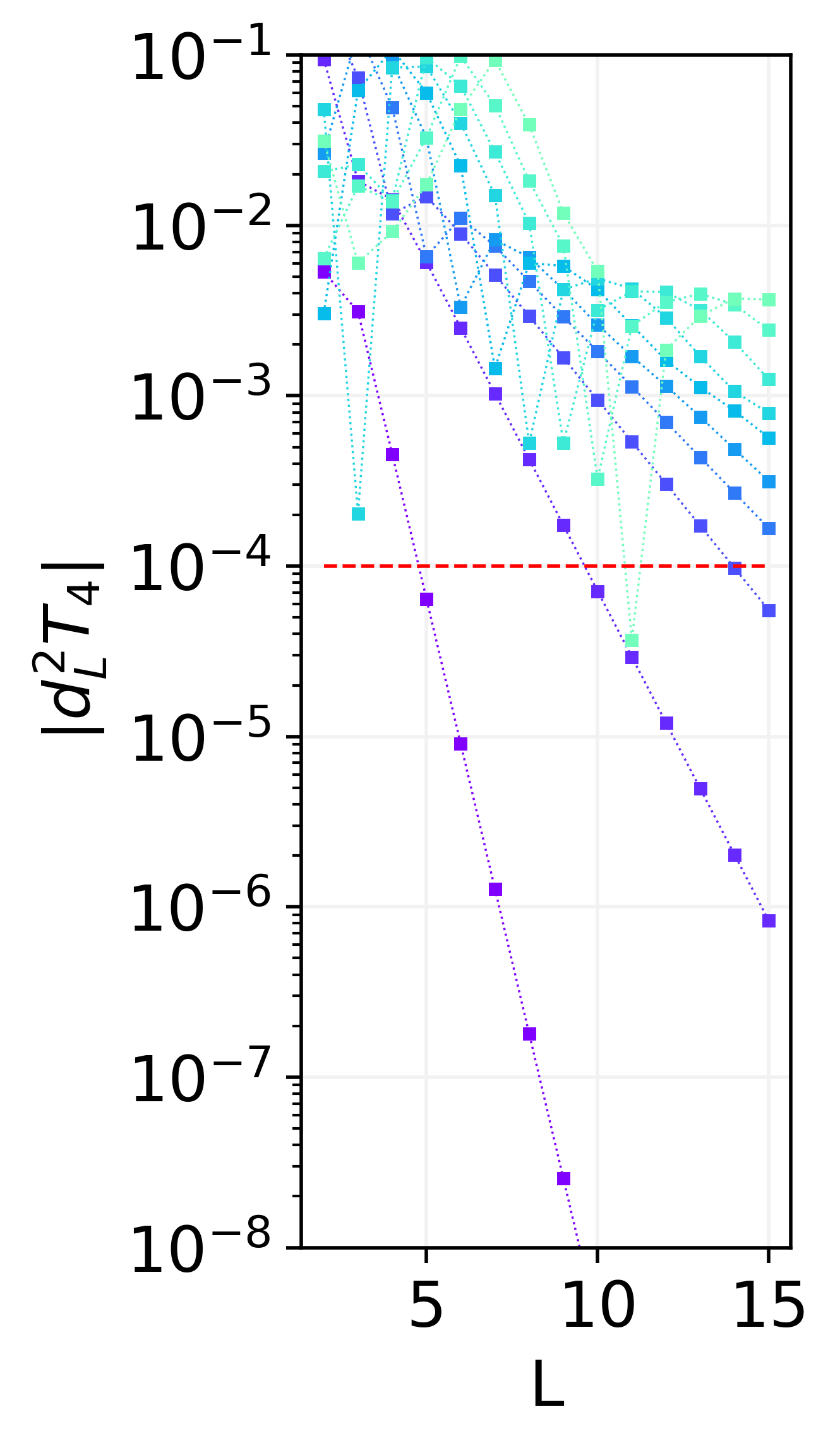}
    \caption{\kern-4em}
  \end{subfigure}\hfill
  \begin{subfigure}[b]{0.25\textwidth}
    \includegraphics[height=65mm]{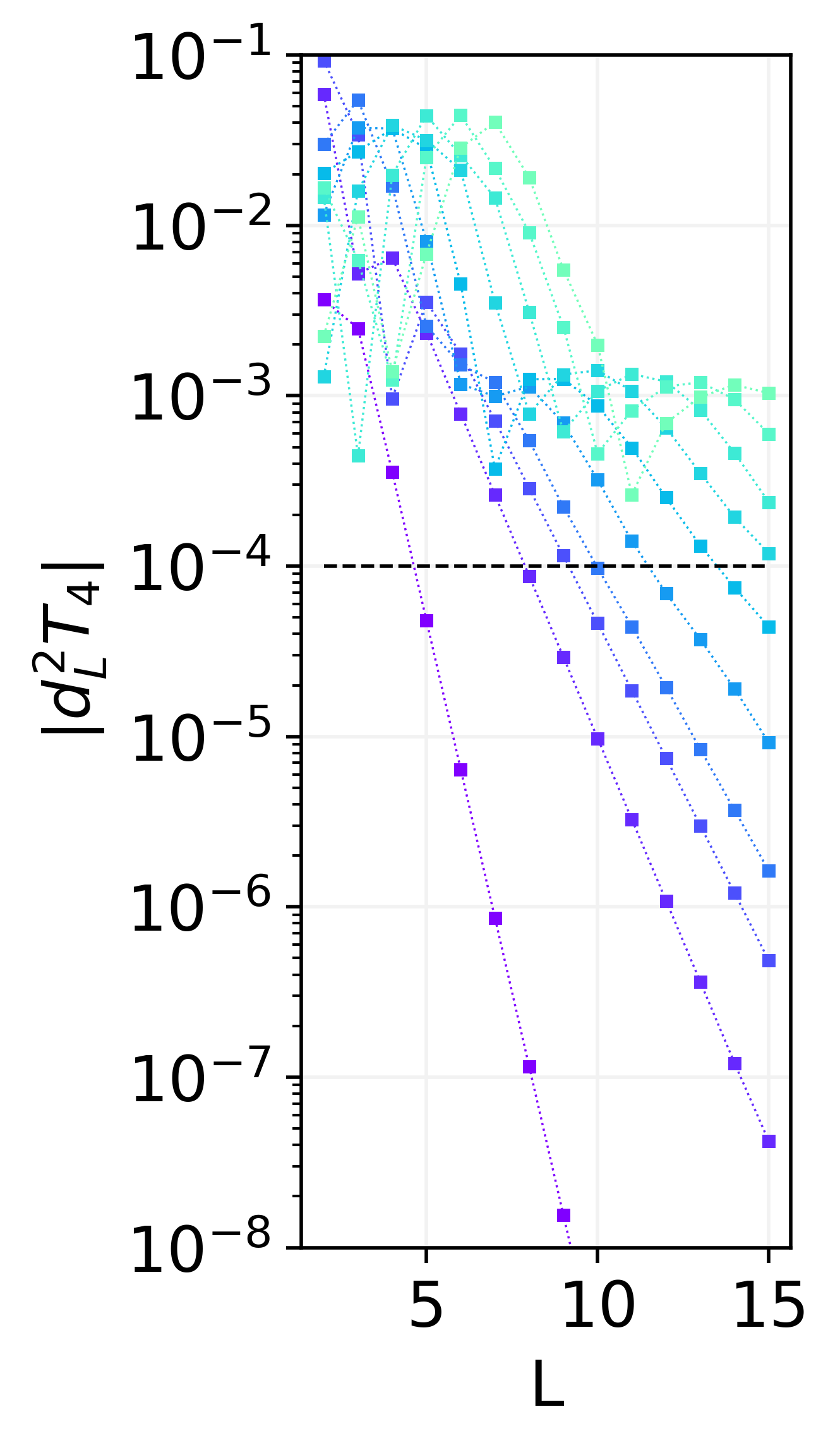}
    \caption{\kern-4em}
  \end{subfigure}
  \caption{\raggedright{Second space derivative of $T_4(\rho_L)$ as a function of the subsystem size $L$ after a quench from $\lambda=10^4$ to different values of $\lambda'$. The subsystem size goes from $2$ to $15$. Different lines represent different times in the range from $0$ (purple line below), to $4$ (red line above). The horizontal black line represents the error tolerance such that functions above the line are definitively exponential. In Panel (a) $\lambda'= 0.5$; in Panel (b) $\lambda'= 1.0$ (critical quench); in Panel (c) $\lambda'= 1.5$; in Panel (d) $\lambda'= 2.0$.} }   \label{fig:ullUpddT4}
\end{figure*}

\begin{figure*}
  \begin{subfigure}[b]{0.3\textwidth}
    \includegraphics[height=65mm]{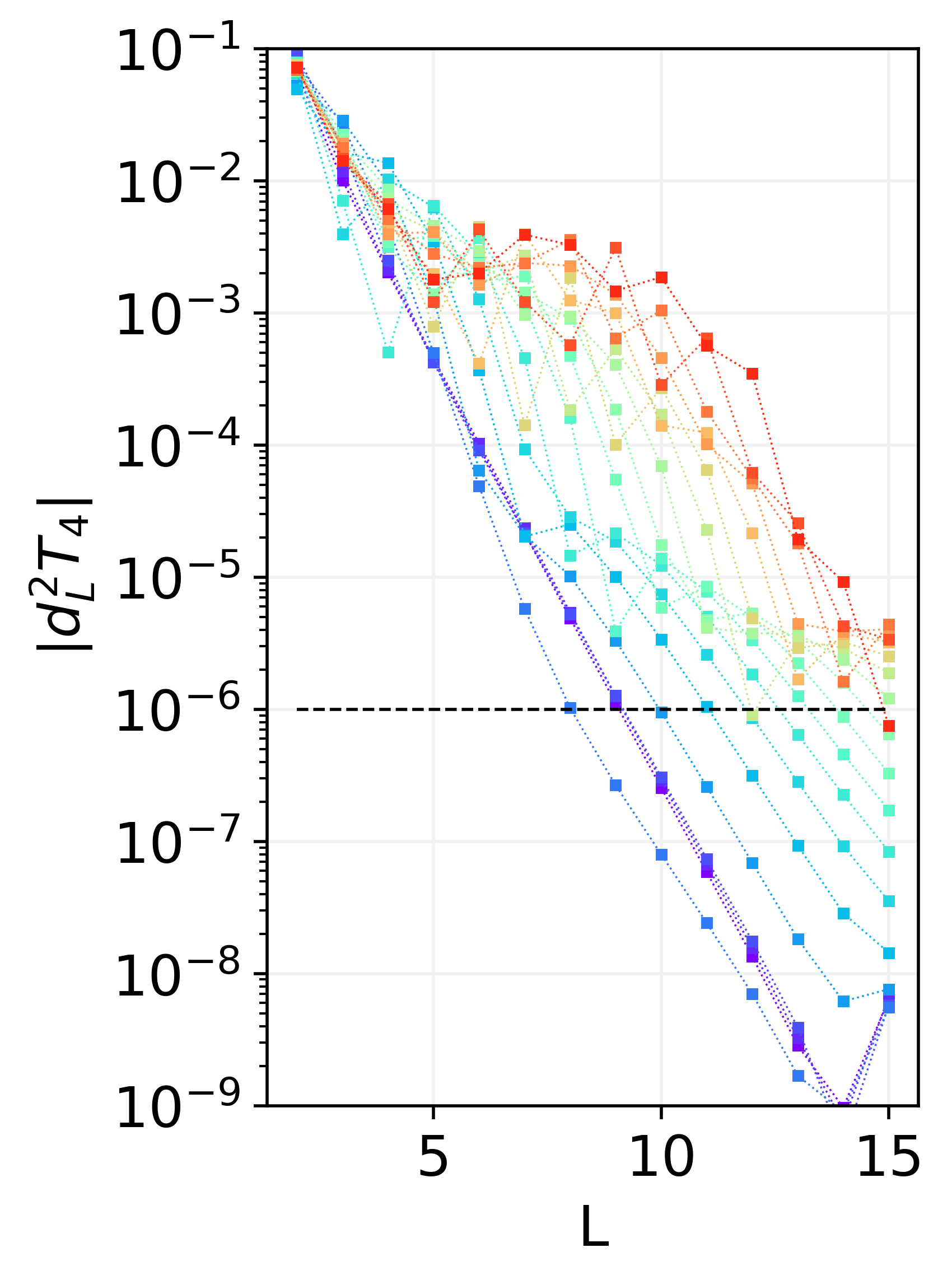}
    \caption{\kern-4em}
  \end{subfigure}\hfill
  \begin{subfigure}[b]{0.3\textwidth}
    \includegraphics[height=65mm]{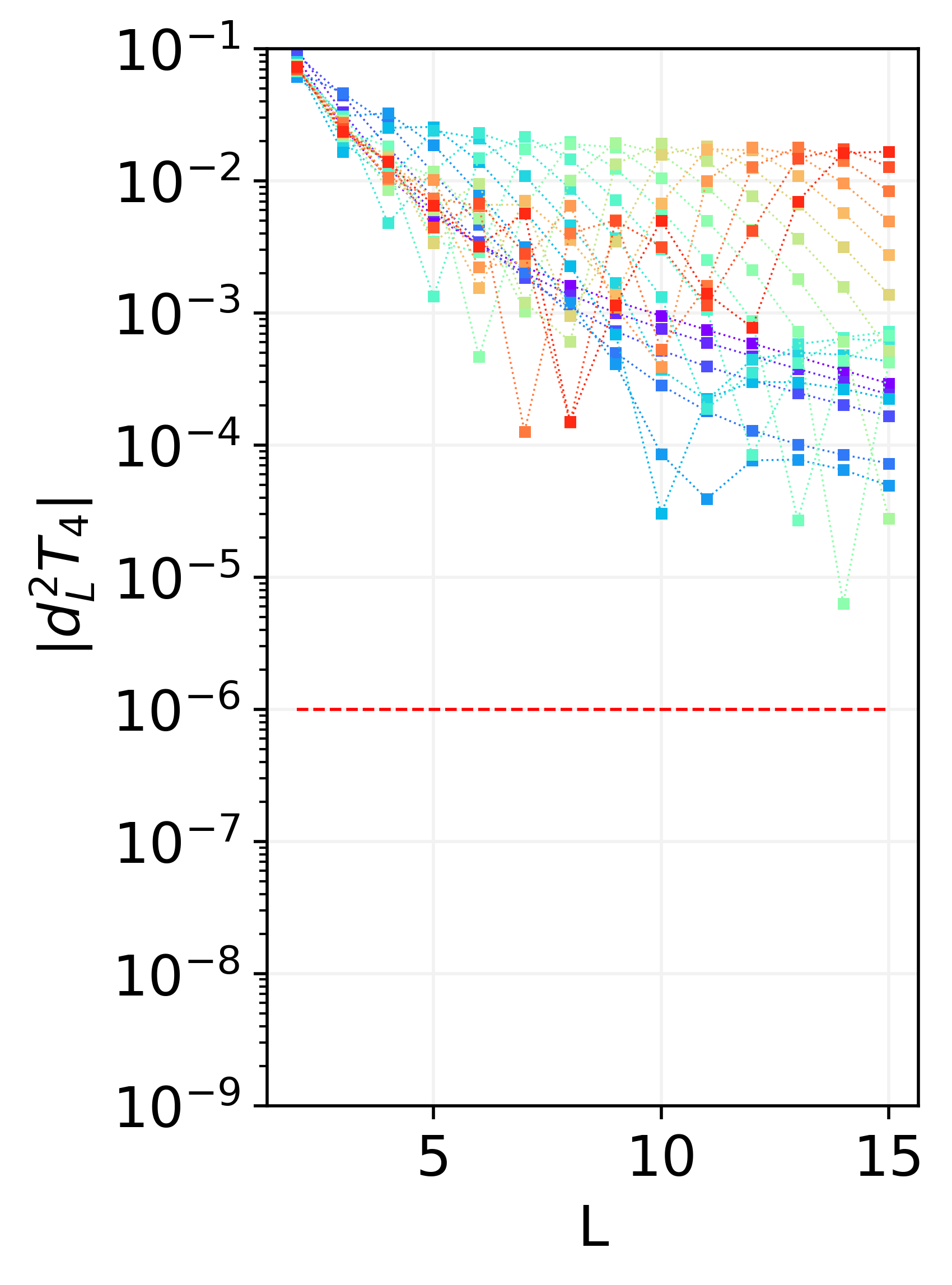}
    \caption{\kern-4em}
  \end{subfigure}\hfill
  \begin{subfigure}[b]{0.3\textwidth}
    \includegraphics[height=65mm]{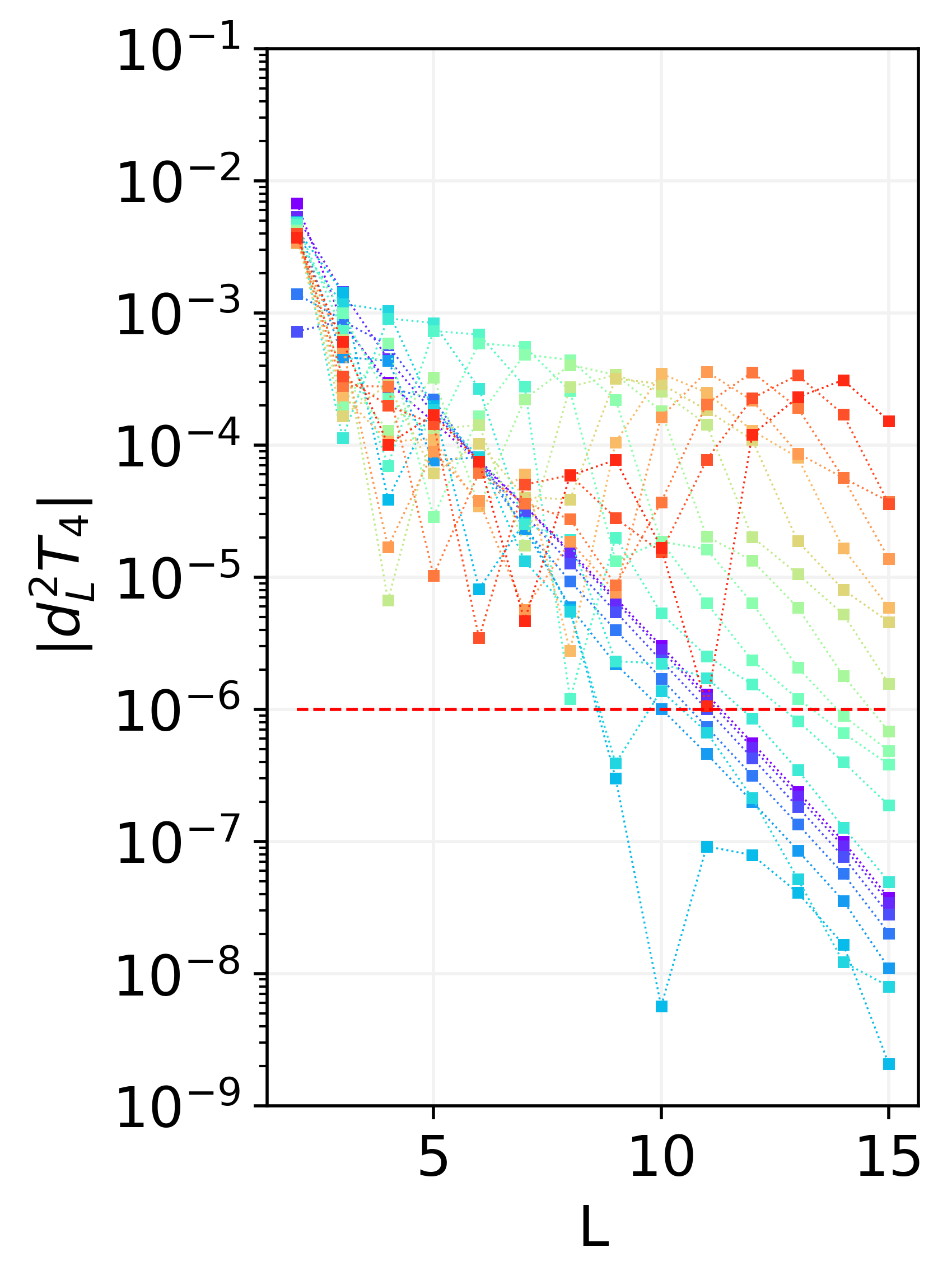}
    \caption{\kern-4em}
  \end{subfigure}
  \caption{\raggedright{Second space derivative of $T_4(\rho_L)$ as a function of the subsystem size $L$ after a quench from $\lambda$ to different values of $\lambda'=\lambda+0.1$. The subsystem size goes from $2$ to $15$. Different lines represent different times in the range from $0$ (purple line below), to $4$ (red line above). The horizontal black line represents the error tolerance such that functions above the line are definitively exponential. In Panel (a) $\lambda= 0.5$; in Panel (b) $\lambda= 0.9$ (critical quench); in Panel (c) $\lambda= 1.5$.}}         \label{fig:smallddT4}
\end{figure*}

In Figures \ref{fig:allUpddT2} and \ref{fig:smallddT2}, we show the exponential decay of the correction to the linear behavior of $T_2$, encoded in the second space derivative, after different quenches. Analogously, in Figures 
\ref{fig:ullUpddT4} and \ref{fig:smallddT4} we represent the exponential decay of the correction to the linear behavior of $T_4$.

In each figure, we can see that the exponential behavior is definitive for a sufficiently large system, as predicted in Section \ref{App:proofofthemainresults}. The size of this large system depends on time and on the quench protocol. Therefore, capturing the definitive behavior is not always possible with our computational resources. When the definitive behavior of the derivative is not captured by our data (e.g. \ref{fig:smallddT2} panel (b)) the dashed line is red, otherwise it is black. In this latter case, we observe that the derivative decays definitively exponentially below the dotted line.

\section{Lieb-Robinson speed}\label{app:lr_speed}

In this Appendix, we analyze the revivals in the Ising model after a quantum quench to reconstruct the associated Lieb-Robinson speed.

\begin{figure*}[th]
  \begin{subfigure}[b]{0.3\textwidth}
    \includegraphics[height=55mm]{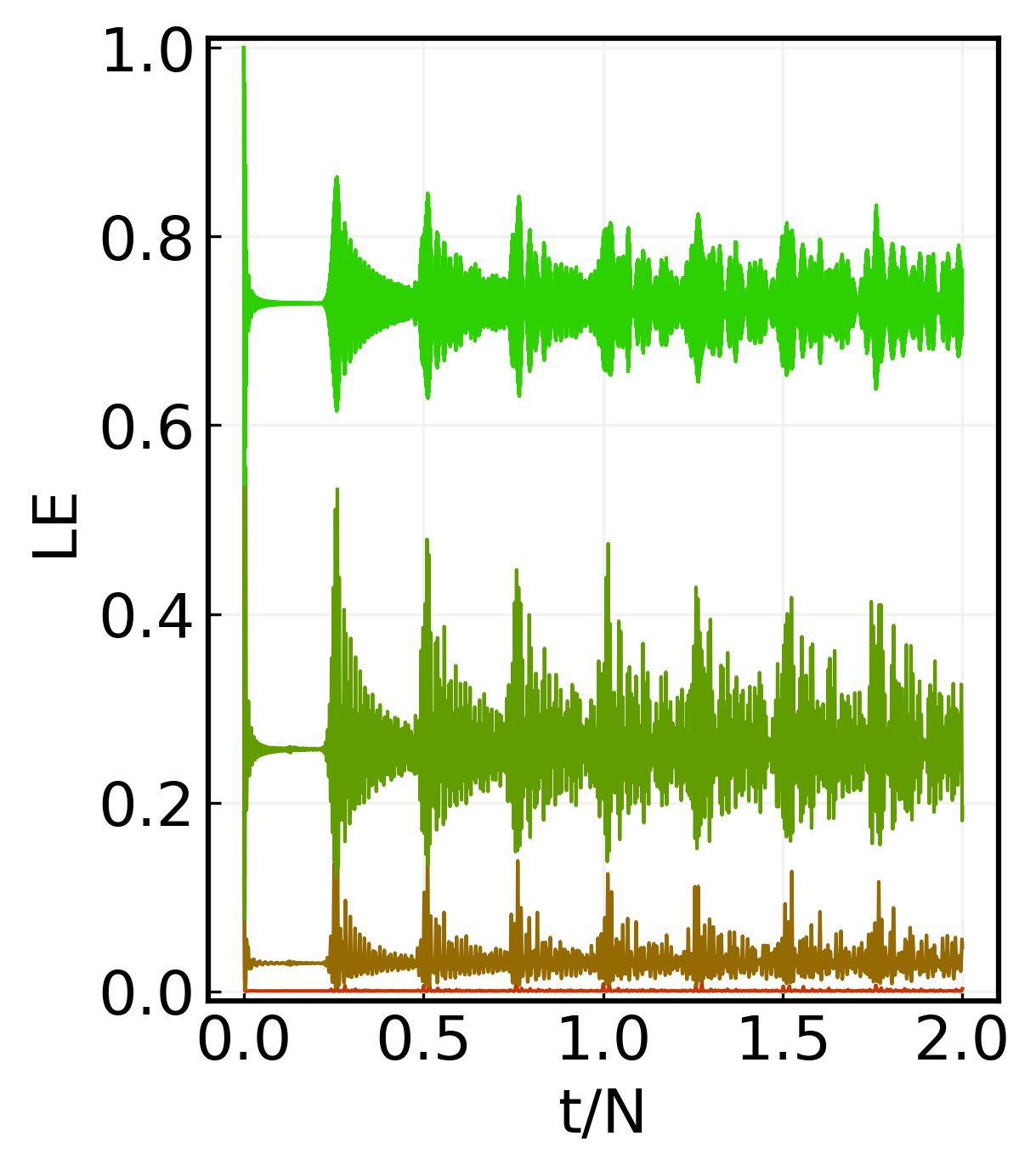}
    \caption{\kern-4em}
  \end{subfigure}\hfill
  \begin{subfigure}[b]{0.3\textwidth}
    \includegraphics[height=57mm]{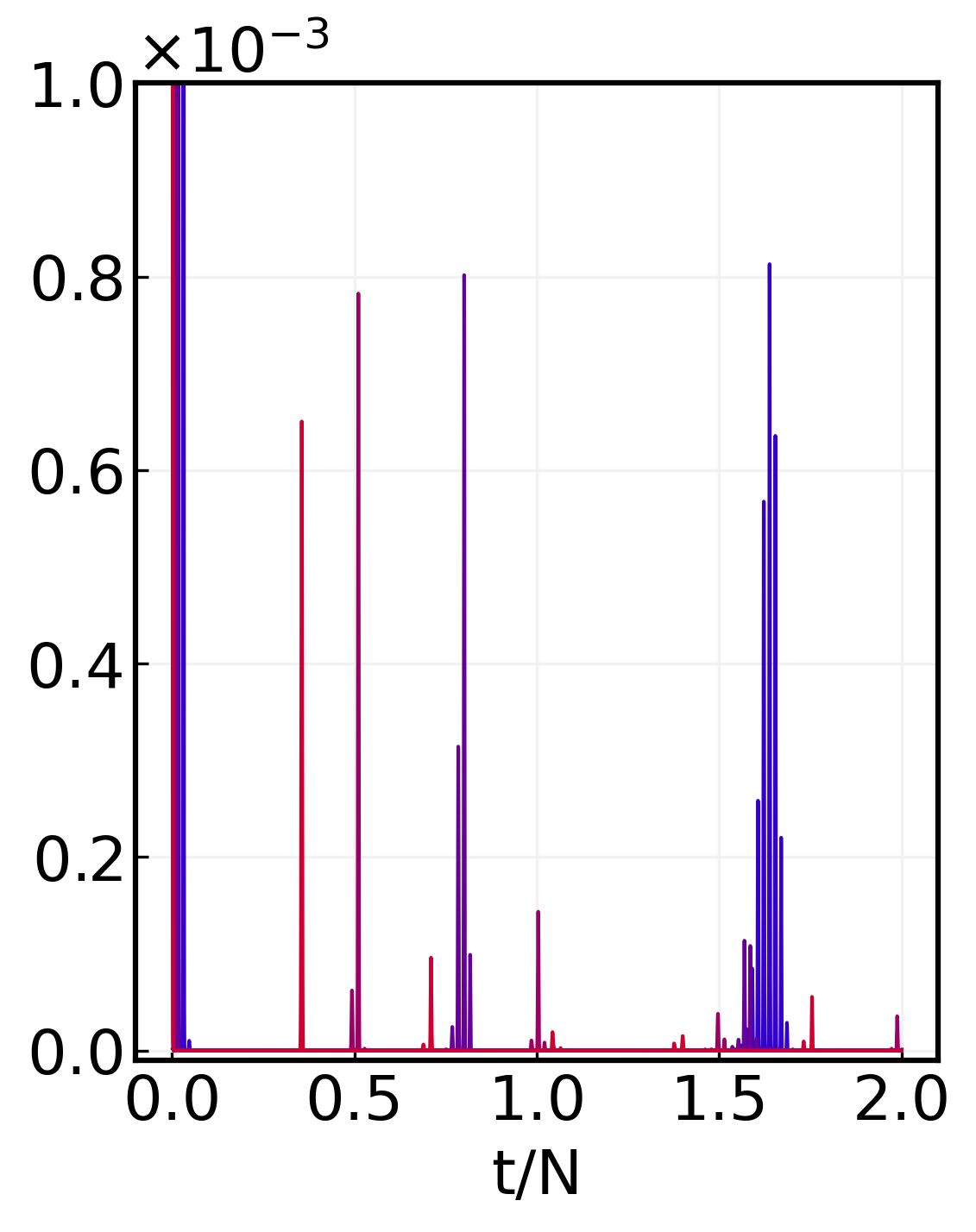}
    \caption{\kern-1.5em}
  \end{subfigure}\hfill
  \begin{subfigure}[b]{0.4\textwidth}
    \includegraphics[height=55mm]{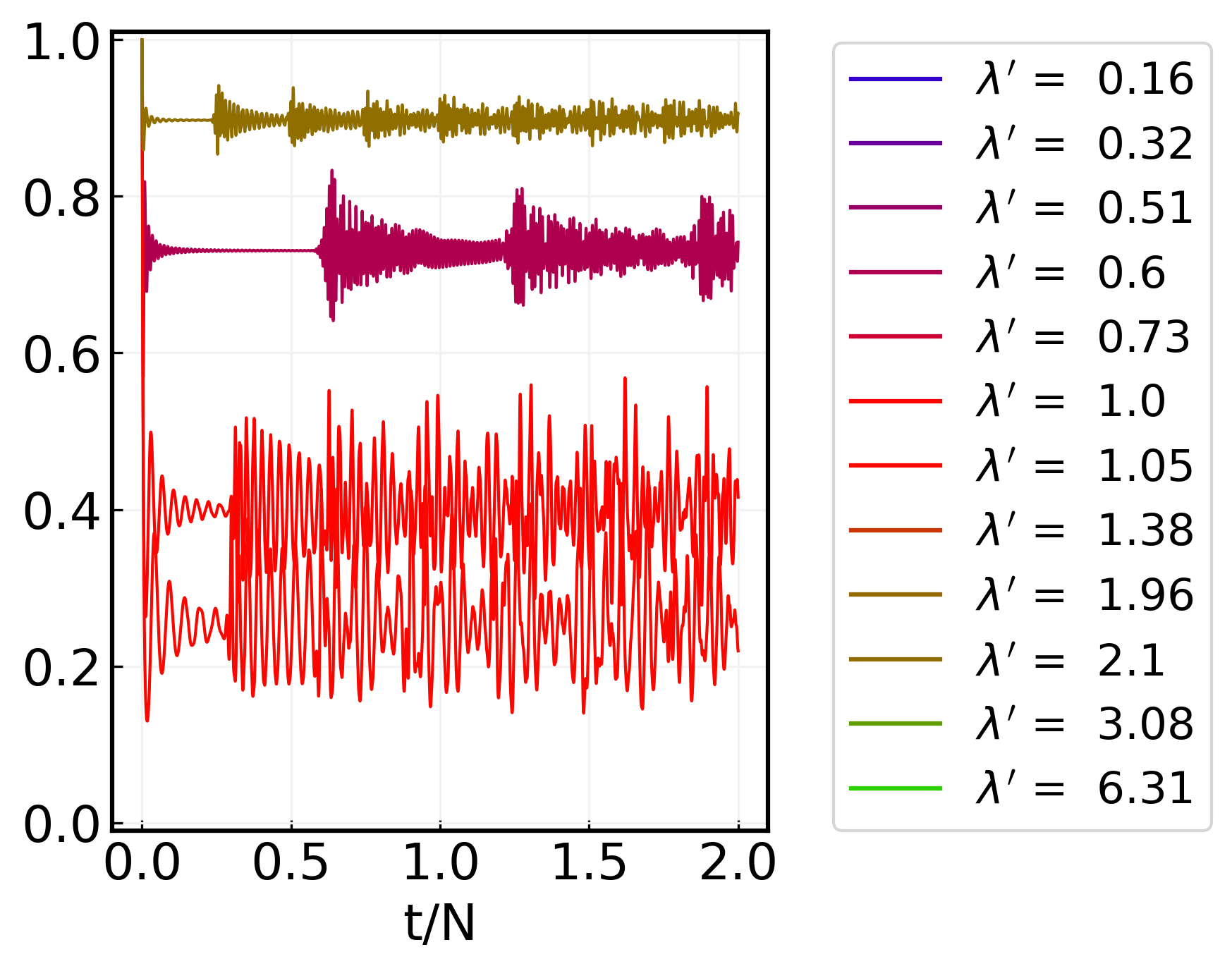}
    \caption{\kern4.5em}
  \end{subfigure}\hfill
  \caption{\raggedright{Evolution of the LE with respect to the rescaled time $t/N$, where $t$ runs from $0$ to $2N$. In Panels (a) and (b) for the large quench protocol where the initial state is $\ket{\psi}\approx\ket{\uparrow\dots\uparrow}$, for a system of $N=100$ spins. In Panel (c) for the small quench protocol where $\lambda'=\lambda+0.1$, for a system of $N=200$ spins.}}
    \label{fig:LE}
\end{figure*}

In Figure \ref{fig:LE} we plot the LE evolution as a function of the rescaled time $t/N$, after the large quench, in Panels (a) and (b), and a large quench, in Panel (c). The revival time is $T_\text{rev}\approx N/4$ for $\lambda'>1$, and $T_\text{rev}\approx N/(\lambda'4)$ for $\lambda'<1$. Therefore, the associated velocity is $v_\text{LR}=2$ for $\lambda'>1$, and $v_\text{LR}=2 \lambda'$ for $\lambda'<1$. As expected from the fact that the Lieb-Robinson velocity only depends on the quench Hamiltonian, the same behavior is encountered both in the small quench and in the large quench.

 \clearpage

%%%%%%%%%%%%%%%%%%%%%%%%%%%%%%%%%%%%%%%%%%%%%%%%%%%%
%BIBLIOGRAPHY
%%%%%%%%%%%%%%%%%%%%%%%%%%%%%%%%%%%%%%%%%%%%%%%%%%%%

%merlin.mbs apSEv4-1.bst 2010-07-25 4.21a (PWD, AO, DPC) hacked

\end{document}